\documentstyle[11pt,aas2pp4]{article}

\lefthead{Brogan et al.}
\righthead{Magnetic Fields Toward M17}

\begin{document}

\newcommand\HII{H\,{\sc ii}}
\newcommand\HI{H\,{\sc i}}
\newcommand\OI{[O\,{\sc i}] 63 $\mu$m}
\newcommand\CII{[C\,{\sc ii}] 158 $\mu$m}
\newcommand\CI{[C\,{\sc i}] 370 $\mu$m}        
\newcommand\SiII{[Si\,{\sc ii}]}
\newcommand\kms{km~s$^{-1}$}
\newcommand\cmt{cm$^{-2}$}
\newcommand\cc{cm$^{-3}$}
\newcommand\CeO{C$^{18}$O}
\newcommand\tCO{$^{12}$CO}
\newcommand\thCO{$^{13}$CO}
\newcommand\Blos{$B_{los}$}
\newcommand\BS{$B_S$}
\newcommand\Bscrit{$B_{S,crit}$}
\newcommand\Bw{$B_W$}
\newcommand\mum{$\mu$m}
\newcommand\muG{$\mu$G}
\newcommand\dv{$\Delta v_{FWHM}$}
\newcommand\va{$v_A$}
\newcommand\Np{$N_p$}
\newcommand\np{$n_p$}
\newcommand\gtsim{\raisebox{-.5ex}{$\;\stackrel{>}{\sim}\;$}}
\newcommand\We{${\cal W}$}
\newcommand\Ms{${\cal M}_S$}
\newcommand\Te{${\cal T}$}

\title{Detection of Magnetic Fields Toward M17 through the  \\
\HI\/ Zeeman Effect}
\author{C. L. Brogan and T. H. Troland}
\affil{University of Kentucky, Department of Physics \& Astronomy, 
Lexington, Ky 40506}
\and
\author{D. A. Roberts, and R. M. Crutcher}
\affil{University of Illinois, Department of Astronomy, Urbana, IL 61801}
\authoremail{brogan@pa.uky.edu}

\begin{abstract}

We have carried out VLA Zeeman observations of \HI\/ absorption lines toward the \HII\/ region
in the M17 giant molecular cloud complex.  The resulting maps have
60\arcsec\/$\times$45\arcsec\/ spatial resolution and 0.64 \kms\/ velocity separation.  The
\HI\/ absorption lines toward M17 show between 5 and 8 distinct velocity components which vary
spatially in a complex manner across the source.  We explore possible physical connections
between these components and the M17 region based on calculations of \HI\/ column densities,
line of sight magnetic field strengths, as well as comparisons with a wide array of previous
optical, infrared, and radio observations.

In particular, an \HI\/ component at the same velocity as the southwestern molecular cloud
(M17~SW) $\sim20$~\kms\/ seems to originate from the edge-on interface between the \HII\/ region
and M17~SW, in un-shocked PDR gas.  We have detected a steep enhancement in the 20 \kms\/ \HI\/
column density and line of sight magnetic field strengths (\Blos\/) toward this boundary.  A
lower limit for the peak 20 \kms\/ \HI\/ column density is
$N_{HI}$/$T_s\geq5.6~\times10^{19}$~cm$^{-2}$\//K while the peak \Blos\/ is $\sim -$450~\muG\/.
In addition, blended components at velocities of 11$-$17~\kms\/ appear to originate from shocked
gas in the PDR between the \HII\/ region and an extension of M17~SW, which partially obscures
the southern bar of the \HII\/ region.  The peak $N_{HI}$/$T_s$ and \Blos\/ for this component
are $\geq4.4 \times10^{19}$~cm$^{-2}$\//K and $\sim+550$~\muG\/, respectively.  Comparison of
the peak magnetic fields detected toward M17 with virial equilibrium calculations suggest that
$\approx 1/3$ of M17~SW's total support comes from its static magnetic field and the other 2/3
from its turbulent kinetic energy which includes support from Alfv\'en waves.

\end{abstract}


\section{INTRODUCTION}

In recent years it has become increasingly clear that magnetic fields play an important role in
the process of star formation.  (See, for example,\markcite{mou76} Mouschovias \& Spitzer
1976; \markcite{hei}Heiles et al.  1993;\markcite{mck93} Mckee et al.  1993).  Unfortunately, the
experimental techniques for measuring the strength and direction of magnetic fields are few and
these methods are observationally challenging.  One such technique uses the Zeeman effect in 21 cm
\HI\/ absorption lines toward galactic \HII\/ regions.  Very Large Array (VLA) observations of the
Zeeman effect in this line yield maps of the line-of-sight magnetic field strengths (\Blos\/).
These maps can then be compared to maps of the distribution and kinematics of ionized, atomic, and
molecular gas to study the role of magnetic fields in star forming regions.  Moreover, estimates
of the masses, column densities, and volume densities of interstellar material associated with the
star-forming regions can be compared to field strengths measured via the Zeeman effect.  These
comparisons are needed to determine the energetic importance of magnetic fields in star-forming
regions.  For studies of this type, the M17 \HII\/ region$-$molecular cloud complex is ideal
because it has been extensively studied at many wavelengths.

The M17 \HII\/ region is one of the strongest thermal radio sources in our galaxy.  Its distance has
been estimated through photometric and kinematic means to be $\sim$ 2.2 kpc (Chini, Els\"{a}sser,
\& Neckel \markcite{chin} 1980;\markcite{reif} Reifenstein et al.  1970;\markcite{wil} Wilson et
al.  1970).  There are at least 100 stars in the M17 \HII\/ region with one O4 V (Kleinmann's star)
and three O5 V stars providing the bulk of the ionizing radiation (\markcite{fel}Felli, Churchwell,
\& Massi 1984 and references therein; see Fig.  1).  The \HII\/ region is made up of two distinct
bar like structures which are $\sim5.7$ pc long and $\sim1.1$ to 1.5 pc wide.  The southern bar has
nearly twice the peak radio continuum flux as the northern bar, and is located just to the east of
the M17~SW molecular cloud core.

The near absence of optical radiation from the southern bar indicates that this region suffers
much more optical extinction than the northern bar (\markcite{gul}Gull \& Balick 1974, Fig 1).
\markcite{dic}Dickel (1968) and \markcite{gat}Gately et al.  (1979) estimate $A_v \sim$ 1-2 mag toward the
northern bar.  Estimates of the visual extinction toward the southern bar range from $A_v \sim
10$ near the radio continuum peak to $A_v\sim 200$ further west toward the core of the adjacent
molecular cloud (M17~SW) (\markcite{beetz}Beetz et al.  1976; \markcite{fel}Felli et al.
1984; \markcite{gat}Gatley et al.  1979; \markcite{thro}Thronson \& Lada 1983).  The low
extinction toward the northern bar seems to indicate that the northern molecular cloud, first
identified by \markcite{lad76}Lada et al.  (1976) ($v_{LSR}\sim 23$ \kms\/), is behind the
\HII\/ region (\markcite{chrys}Chrysostomou et al.  1992).

The most interesting characteristic of M17 is that the interface between the southern bar of the
\HII\/ region and M17~SW is seen almost edge on (Fig.  1).  From models of the density gradient on
the western edge of the \HII\/ region, \markcite{ick}Icke, Gatley, \& Israel (1980) estimated that
this interface lies at an angle of $\sim 20\arcdeg$ with respect to the line of sight.  Although the
transition from the \HII\/ region to the molecular cloud is quite sharp, the interface region is too
wide in photodissociation region (PDR) tracers such as \CII\/ unless it is clumpy (see eg.,
\markcite{stut88}Stutzki et al.  1988).  The extended \CII\/ emission observed in M17~SW by
\markcite{stut88}Stutzki et al.  (1988) and \markcite{bor}Boreiko, Betz, \& Zmuidzinas (1990)
requires 912-2000\AA~UV photons to persist into the molecular cloud at least an order of magnitude
further than the predicted UV absorption length scale.  That is, atoms are being photoionized and
molecules photodissociated further into the molecular cloud than one would expect for a homogeneous
medium.  (See \markcite{tei}Tielens \& Hollenbach 1985 for homogeneous model calculations.)  A
possible explanation for this paradox is that the interface region is clumpy.

Evidence that M17 SW is clumpy has been observed with high angular resolution at many
\newpage
\onecolumn
\begin{figure}
\vspace{-4.0cm}\hspace{1.0cm}
\epsfxsize=13.0cm \epsfbox{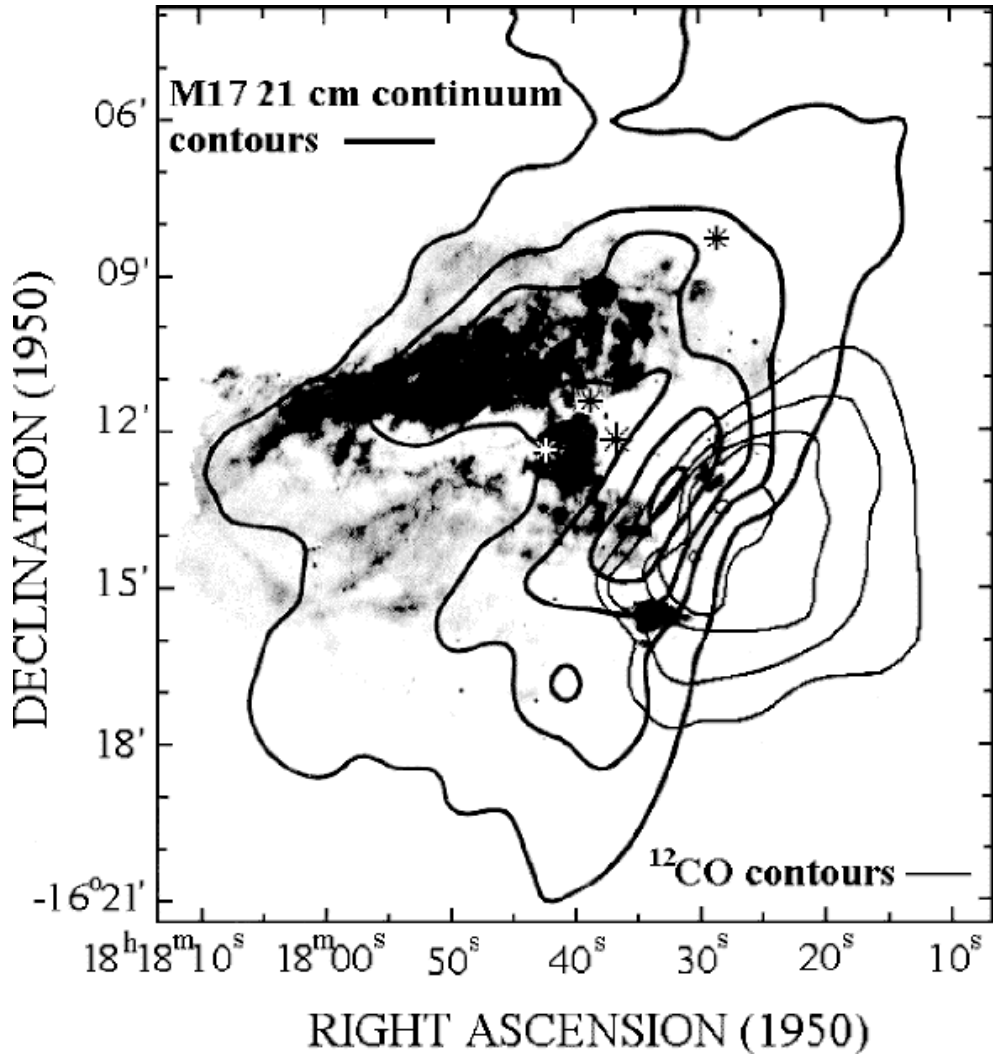}
\vspace{-.0cm}
\caption[fig1.img2.eps]{Map of the 1, 5, 10, 20, and 30 Jy beam$^{-1}$ 21 cm continuum contours
(present work, ({\em thick lines}) superimposed on an optical image from Gull \& Balick
and simplified \tCO\/ contours ({\em thin lines}) from Thronson \& Lada
(1983).  The four stars which are primarily responsible for ionizing the \HII\/
region are indicated with star symbols, with the largest representing Kleinmann's star (Felli,
Churchwell, \& Massi 1984) \label{figu1}.}
\end{figure}
\twocolumn
\noindent wavelengths.  For example, \markcite{fel}Felli et al.  (1984), found clumpiness in their $\sim
10\arcsec$ resolution radio continuum observations of the M17 \HII\/ region.  They suggest this
clumpiness originates from high density neutral clumps surrounded by dense ionized envelopes.
These clumps may be the remnants of clumps from the parent molecular cloud which have been
overtaken by the \HII\/ region ionization front.  \markcite{mas}Massi, Churchwell, \& Felli
(1988) observed seven clumps in NH$_3$ emission with 6\arcsec\/ resolution toward the
northwestern portion of the \HII\/ region$-$M17 SW interface.  The average density and size of
these clumps are $n_{H_2} \sim 10^{7}$~\cc\/ and 0.05 pc.  \markcite{stut90}Stutzki \&
G\"{u}sten (1990) modeled their $\sim 13\arcsec$ resolution \CeO\/ ($2\rightarrow 1$)
observation (with a larger field of view than the NH$_3$ observation) with $\sim 180$ clumps
with average densities of $n_{H_2}$ $\sim 10^{5-6}$~\cc\/ and diameters of $\sim$ 0.1~pc.
\markcite{hob92}Hobson (1992) also saw evidence for clumpiness in HCO$^{+}$ and HCN molecules
at $\sim 19\arcsec$ resolution, and \markcite{wan}Wang et al.  (1993) found clumps consistent
with the \CeO\/ clumps in $\sim 18\arcsec$ to $\sim 34\arcsec$ resolution observations of CS and
C$^{34}$S molecules.

The identification of clumps in M17~SW is significant for several reasons.  For example,
calculations of density are more complex due to its dependence on the optical depth of the
observed species, and the assumed clump filling factor.  As discussed previously, the
lengthscale of the PDR also depends on the clumpiness of the region.  Therefore, the regions and
morphology of photodissociated atomic gas are difficult to predict.  Also the random motion of
such clumps can influence observed linewidths and the sizescale of turbulence within the
molecular cloud.  These considerations play an important role in our analysis of the \HI\/ gas
and the line of sight magnetic fields calculated from it.

In this paper we report the characteristics of the 21 cm continuum (\S 3.1), \HI\/ optical
depths and column densities (\S 3.2), and the line of sight magnetic field strengths (\S 3.3)
derived from our VLA \HI\/ Zeeman effect observation.  We discuss possible origins for the most
prominent \HI\/ absorption components in \S 4.1.  In addition, we present comparisons of our 20
\kms\/ \Blos\/ map with previous linear polarimetry observations in \S 4.2, and we also compare
expected magnetic field strengths from virial equilibrium arguments to our observed line of
sight magnetic field strengths in \S 4.3.  Our findings are summarized in \S 5.

\section{OBSERVATIONS} 

\placetable{tab1}

Our \HI\/ absorption data were obtained with the DnC configuration of the VLA \footnote{The
National Radio Astronomy Observatory is operated by Associated Universities, Inc., under
contract with the National Science Foundation.}.  Key parameters from this observation can
be found in Table 1.  We observed both senses of circular polarization simultaneously.
Since the Zeeman effect is very sensitive to small variations in the bandpass, we switched
the sense of circular polarization passing through each telescope's IF system every 10
minutes with a front-end transfer switch.  In addition, we observed each of the calibration
sources at frequencies shifted by $\pm$ 1.2 MHz from the \HI\/ rest frequency to avoid
contamination from Galactic \HI\/ emission at velocities near those of M17.

The calibration, map making, cleaning, and calculation of optical depths were all carried
out using the AIPS (Astronomical Image Processing System) package of the NRAO.  The right
(RCP) and left (LCP) circular polarization data were separately calibrated and mapped using
natural weighting.  These two data cubes were then combined to create Stokes I (RCP + LCP)
and Stokes V (RCP $-$ LCP) cubes.  The I and V cubes were CLEANed using the AIPS task SDCLN
down to 70 mJy beam$^{-1}$.  Bandpass correction was applied only to the I data since bandpass
effects subtract out to first order in the V data.  Subsequent antenna leakage correction and
magnetic field derivations were carried out using the MIRIAD (Multichannel Image
Reconstruction Image Analysis and Display) processing package from BIMA.

\section{RESULTS}

\subsection{The 21 cm Continuum}

\placefigure{figu1} 

Figure~1 shows our 21 cm continuum contours with an optical image from Gull \& Balick
\markcite{gul} (1974) and simplified \tCO\/ contours from \markcite{thro}Thronson \& Lada (1983)
superimposed.  The total continuum flux at 21 cm is $\sim800$ Jy calculated inside the 1 Jy
beam$^{-1}$ (250 K) contour level (lowest contour in Fig.  1).  This value is almost twice that
obtained by Lockhart \& Goss \markcite{loc} (1978) with a 2\arcmin\/ beam at the Owens Valley
interferometer.  In addition, we estimate the total flux of the Southern bar is 384 Jy while that
of the Northern bar is 240 Jy (calculated for points inside the 5 Jy beam$^{-1}$ contour level).
Our total flux inside the 5 Jy beam$^{-1}$ contour (624 Jy) is similar to the 644 Jy reported by
Lada et al.  \markcite{lad76} (1976) with the Haystack 120 ft telescope and 620 Jy observed by
L\"{o}bert \& Goss \markcite{lob} (1978) with the Fleurs synthesis telescope at 21 cm.

\subsection{Atomic Hydrogen Optical Depths and Column Densities}

\begin{figure}
\vspace{-2.4cm}\hspace{2.0cm}
\epsfxsize=9.5cm \epsfbox{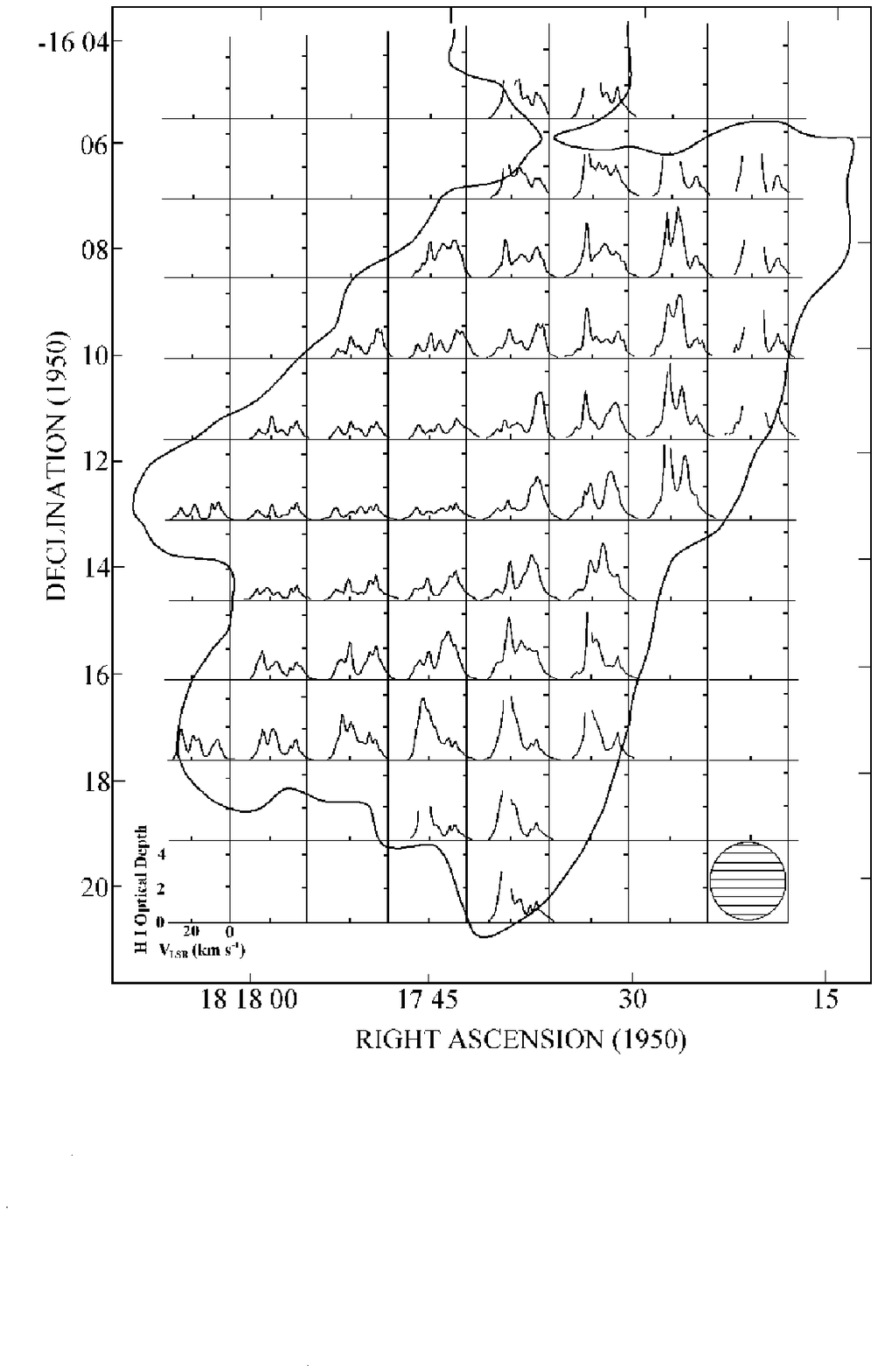}
\vspace{-4.9cm}
\figcaption[cryopt15.eps]{\HI\/ Optical depth profiles across the source after convolution with a
2\arcmin\/ beam (each profile is independent).  The 1 Jy beam$^{-1}$ 21 cm continuum contour,
and $2\arcmin$ beam are shown for reference \label{figu2}.}
\vspace{0.1cm}\hspace{2.0cm}
\epsfxsize=9.0cm \epsfbox{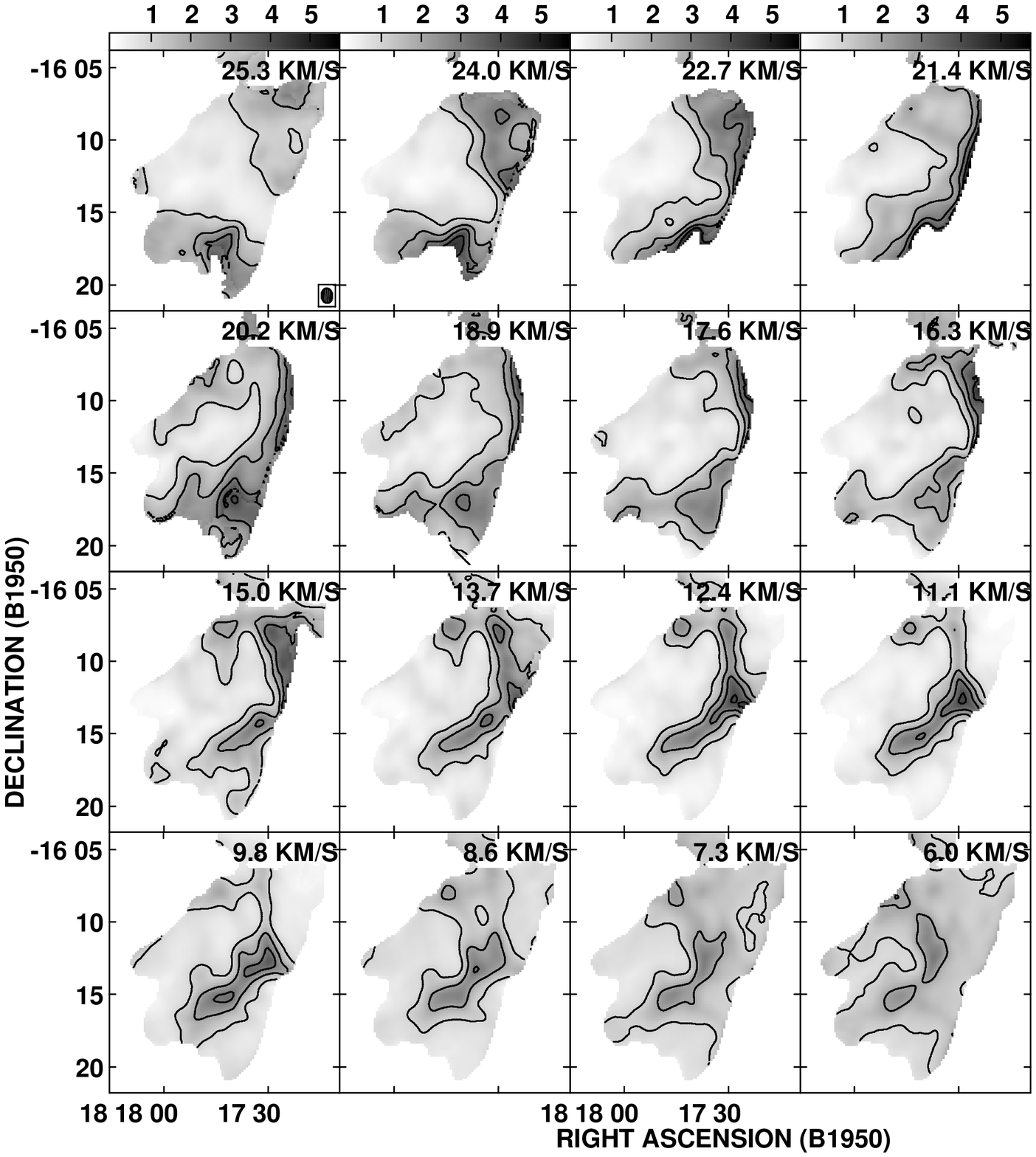}
\caption[CHAN2AVG.PS]{Velocity channel maps of the \HI\/ optical depth morphology (after
averaging every two channels) from 6.0 to 25.3 \kms\/ (greyscale), with contours at $\tau=$1, 2,
3, 4, and 5.  Notice the increase in \HI\/ optical depths as you approach M17 SW in the 24.0 to
17.6 \kms\/ velocity range \label{figu3}.}
\end{figure}

\placefigure{figu2}

M17 \HI\/ optical depths were calculated assuming that the spin temperature $T_s$ is much less
than the background continuum temperature $T_c$ of the \HII\/ region.  In addition, optical
depth calculations were restricted to positions where the continuum power is greater than 1 Jy
beam$^{-1}$ (250 K) and the signal-to-noise ratio is higher than three.  The complex nature of
the absorption lines in M17 can be seen from the optical depth spectra displayed in Figure 2.
The strengths of some components vary spatially over the source but remain distinct, while other
components seem limited to particular regions.  This complexity makes Gaussian fitting very
difficult.  Our profiles agree qualitatively with the Gaussian fits reported by \markcite{loc}
Lockhart \& Goss (1978) made with a $2\arcmin$ beam and 0.85 \kms\/ velocity resolution.  They
fit their M17 \HI\/ optical depth profiles with eight Gaussians having center velocities ranging
from 4.2 to 27.5 \kms\/.  In the following analysis we concentrate on the \HI\/ velocity
component near 20 \kms\/ and the blended components between 11$-$17 \kms\/ due to their high
optical depths and their coincidence in velocity with other species observed in M17 SW (see \S
4.1 for details).

\placefigure{figu3}

Figure 3 displays the morphology of the \HI\/ optical depths toward the M17 \HII\/ region as a
function of velocity (every two adjacent channels were averaged).  The steep increase of the
\HI\/ optical depth toward the \HII\/ region$-$molecular cloud interface is clearly apparent.
This effect is particularly noticeable near 20 \kms\/.  Although the \HI\/ optical depths become
saturated at the western boundary of the \HII\/ region (particularly in the northwest), they
reach values at least as high as 5.  Optical depth maps in the 11$-$17 \kms\/ range have their
highest values further to the east and in the northern part of the source.

\begin{figure}
\vspace{-3.0cm}
\epsfxsize=16.0cm \epsfbox{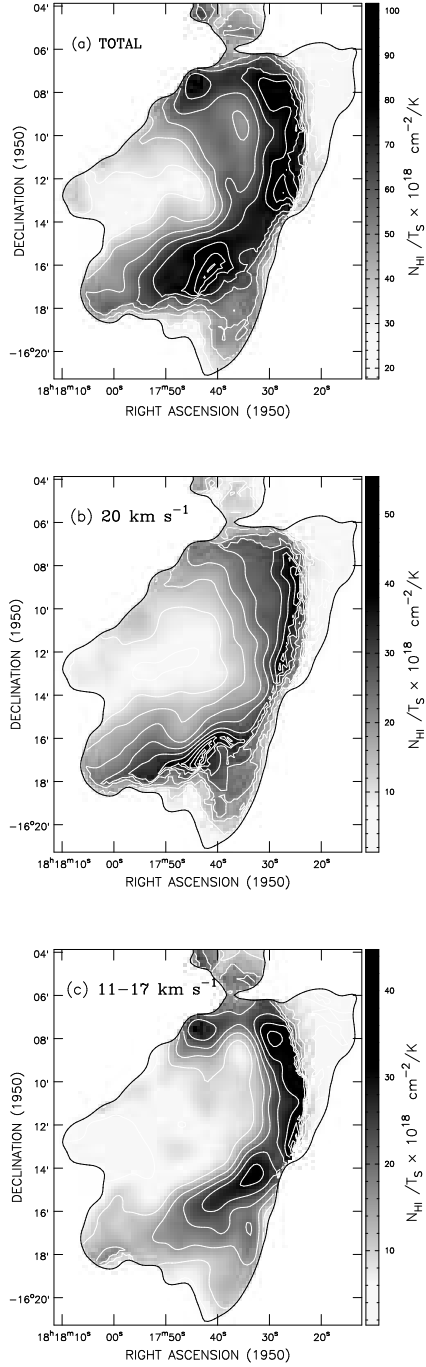}
\caption[FigCLMN.eps]{Maps of \HI\/ column density divided by spin temperature ($N_{HI}/T_s$)
calculated using equation (1).  For each map the outermost black contour is the 1 Jy beam$^{-1}$
21 cm continuum level, and the greyscale and white contours represent $N_{HI}/T_s \times
10^{18}$ cm$^{-2}$/K.  (a) Data summed across the $entire$ \HI\/ line.  White contour levels are
in increments of $10 \times 10^{18}$ cm$^{-2}$/K from 20 to 100 $\times 10^{18}$ cm$^{-2}$/K,
(b) Data for the 20 \kms\/ component summed from 17 to 24.5 \kms\/.  White contour levels are in
increments of 5 $\times 10^{18}$ cm$^{-2}$/K from 5 to 55 $\times 10^{18}$ cm$^{-2}$/K, and (c)
Data for the 11 to 17 \kms\/ component summed from 10.5 to 16.9 \kms\/.  White contour levels
are in increments of 5 $\times 10^{18}$ cm$^{-2}$/K from 5 to 40 $\times 10^{18}$
cm$^{-2}$/K \label{figu4}.}
\end{figure}

\placefigure{figu4}

The column density of \HI\/ toward M17 was found using the relationship 
\begin{equation} 
N_{HI} = 1.823\times10^{18}T_s\int\tau_{\nu}d\nu~{\rm cm^{-2}}, 
\end{equation}
where $\tau_{\nu}$ is the optical depth per unit frequency.  Values for $N_{HI}$/$T_s$ summed
across the entire \HI\/ velocity range, the 20 \kms\/ component (from 17 to 24.5 \kms\/) and the
11$-$17 \kms\/ blended component are displayed in Figs.  4a, 4b, and 4c, respectively.  At
positions along the northwest and southwest portions of the source, these maps represent lower
limits to $N_{HI}$/$T_s$ owing to saturation effects.  Note the increase in the \HI\/ column
density toward the \HII\/ region$-$molecular cloud interface on the southwestern side of the
maps.  There is also spatial agreement between the \HI\/ column density concentration prominent
in the total $N_{HI}$/$T_s$ map (Fig.  4a) near $18{\rm^h} 17{\rm^m} 30.0{\rm^s}$, ${-16}\arcdeg
13\arcmin 05\arcsec$ and the northern condensation seen 
\newpage
\onecolumn
\newpage
\twocolumn
\newpage
\noindent in molecular gas (see \markcite{wan} Wang et al.  1993 and Fig.~8). The total $N_{HI}/T_s$ 
at this position is 8.3 $\times 10^{19}$ \cmt\//K.

Aside from line saturation, the main source of uncertainty in calculating \HI\/ column densities
is in the value assumed for $T_s$.  Upper limits to $T_s$ come from two sources.  For one, the
\HI\/ line widths in some cases are quite narrow.  Toward the southern end of the molecular
ridge defined by CS emission (\markcite{wan}Wang et al.  1993) they are as little as 2.8 \kms\/
for the 20 \kms\/ component (Figs.  8, 9).  This width implies a maximum temperature of 180 K at
this position.  Also, along the periphery of the continuum source, \HI\/ absorption lines exist
where the continuum brightness is as little as 250 K (1 Jy beam$^{-1}$).  Evidently, $T_s$ is
less than about 200 K in many directions toward M17.  At the position illustrated in Fig.  9,
the CS line has a width of about 2 \kms\/.  If \HI\/ and CS are mixed and at a common
temperature, then $T_s\approx $100 K.  In fact, modeling of IRAS 12 \mum\/ infrared data
by\markcite{hob94}Hobson \& Ward-Thompson (1994) require a hot dust component of $\sim$108 K to
explain the overall 12 \mum\/ emission from M17.  This hot dust can be linked to the PDR by the
fact that warm emission at 12 \mum\/ is likely to arise from a population of polycyclic aromatic
hydrocarbons (PAHs) (\markcite{des}D\'esert, Boulanger \& Puget 1990), and PAH emission from
PDR regions has been predicted by many authors (e.g.  \markcite{holl93}Hollenbach 1993).

Another indication of temperature in the atomic gas comes from the observations of \CI\/ by
\markcite{gen}Genzel et al.  (1988).  They estimate a temperature of 50 K for \CI\/ gas near 20
\kms\/.  Molecular temperature determinations may also be relevant if some of the \HI\/ is mixed
with molecular gas (\S 4.2).  \markcite{bergin}Bergin et al.  (1994) estimate $T_K$ = 50 K from
observations of low-$J$ CO transitions, while \markcite{har}Harris et al.  (1987) show that
high $J$ CO lines ($7\rightarrow6, 14\rightarrow13$) arise from PDR gas at about 250 K.
\markcite{mei}Meixner et al.  (1992) found that the spatial extent and intensities of a wide
array of far-infrared and submillimeter cooling lines could be adequately modeled by a three
component gas consisting of very dense clumps $n\sim 5\times 10^5$~\cc\/, intermediate density
interclump gas $n\sim 5\times 10^3$~\cc\/, and a tenuous extended halo $n\sim 3\times
10^2$~\cc\/.  They obtained the best agreement with observations by assuming an interclump gas
kinetic temperature of $\sim 200$ K where it is also likely that dissociated \HI\/ gas can be
found.  Almost certainly, $T_s$ in the \HI\/ gas is quite variable, with values in the range 50
- 200 K most common.

As described above, it is likely that no one spin temperature is applicable to all of the \HI\/
gas components seen in M17.  With a spin temperature of 50 K, a lower limit to the peak $total$
column density toward M17 is $N_{HI}=5.1\times 10^{21}$~\cmt\/.  Spin temperatures between 50 K
and 200 K yield peak column densities of $2.8-11.0\times 10^{21}$~\cmt\/ for the 20 \kms\/
component and $2.2-9.0\times 10^{21}$~\cmt\/ for the 11$-$17 \kms\/ blended components.  We can
use such estimates of the \HI\/ column density to estimate what fraction of the total hydrogen
column density arises from atomic gas.  For example, \markcite{fel}Felli et al.  (1984) used
the optical depth of the silicon absorption feature at 10 \mum\/ to estimate that $A_{\nu}$=30
toward the ultra compact \HII\/ region M17-UC1 ($18{\rm^h}17{\rm^m}32{\rm^s},
{-16}\arcdeg13\arcmin00\arcsec$).  Using $N_{H_{TOT}}= 2 \times 10^{21} A_v$ (\markcite{bert92}
Bertoldi\& Mckee 1992), we estimate that the total column density of hydrogen is
$N_{H_{TOT}}\approx 6 \times 10^{22}$~\cmt\/.  Our {\em total} \HI\/ column density toward
M17-UC1 is $N_{HI}\geq 3.7\times 10^{21}$~\cmt\/ ($T_s$=50 K).  Unless saturation effects have
resulted in a severe underestimate of $N_{HI}$, much of the gas along the line of sight to
M17$-$UC1 is in the form of H$_2$.

\subsection{Magnetic Field Strengths From the Zeeman Effect}

The \HI\/ Zeeman effect is only sensitive to the line of sight magnetic field (\Blos\/) in most
astrophysical situations since the Zeeman splitting is a tiny fraction of the line width.  Since
V$\propto d$I$/d\nu$, values for \Blos\/ were obtained by fitting the derivative of the I
profile to the V profile at each pixel with a least squares fitting routine (for details see
Roberts et al.  1993, \markcite{cru96}Crutcher et al.  1996).  Due to the complexity of the
\HI\/ spectra, the number of channels fit for each velocity component is small.  To obtain more
realistic error estimates, forty channels outside of the line region were included in each fit.
The resulting \Blos\/ maps were subsequently masked where the continuum power is less than 1 Jy
beam$^{-1}$ and when \Blos\//$\sigma$(\Blos\/) $<$ 3.  The quantity \Blos\//$\sigma$(\Blos\/)
will subsequently be denoted $S/N_B$.

\subsubsection{Magnetic Fields Derived for the 20 \kms\/ Component} 

\begin{figure}
\vspace{-4.1cm}\hspace{2.5cm}
\epsfxsize=10.0cm \epsfbox{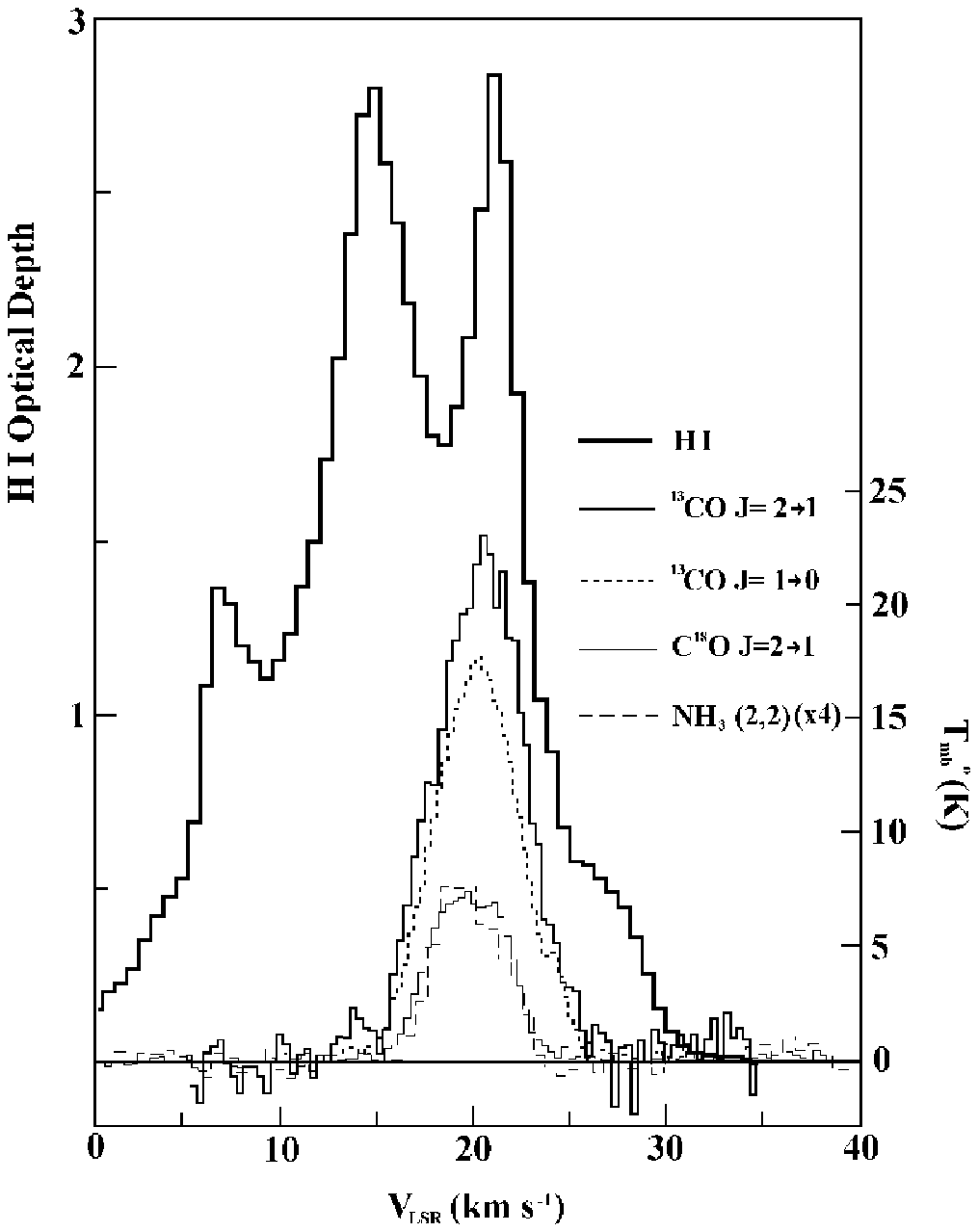}
\vspace{-1.0cm}
\caption[crymulti.eps]{An \HI\/ optical depth profile (present work) and \thCO\/, \CeO\/, and
NH$_3$ emission profiles from Stutzki et al.  1988 toward the position
$18^{\rm h}17^{\rm m}30.6^{\rm s}$, ${-16}\arcdeg14\arcmin00\arcsec$ (also known as position
(-60,-30) in Stutzki et al.) \label{figu5}.}
\vspace{1.0cm}\hspace{2.0cm}
\epsfxsize=11.0cm \epsfbox{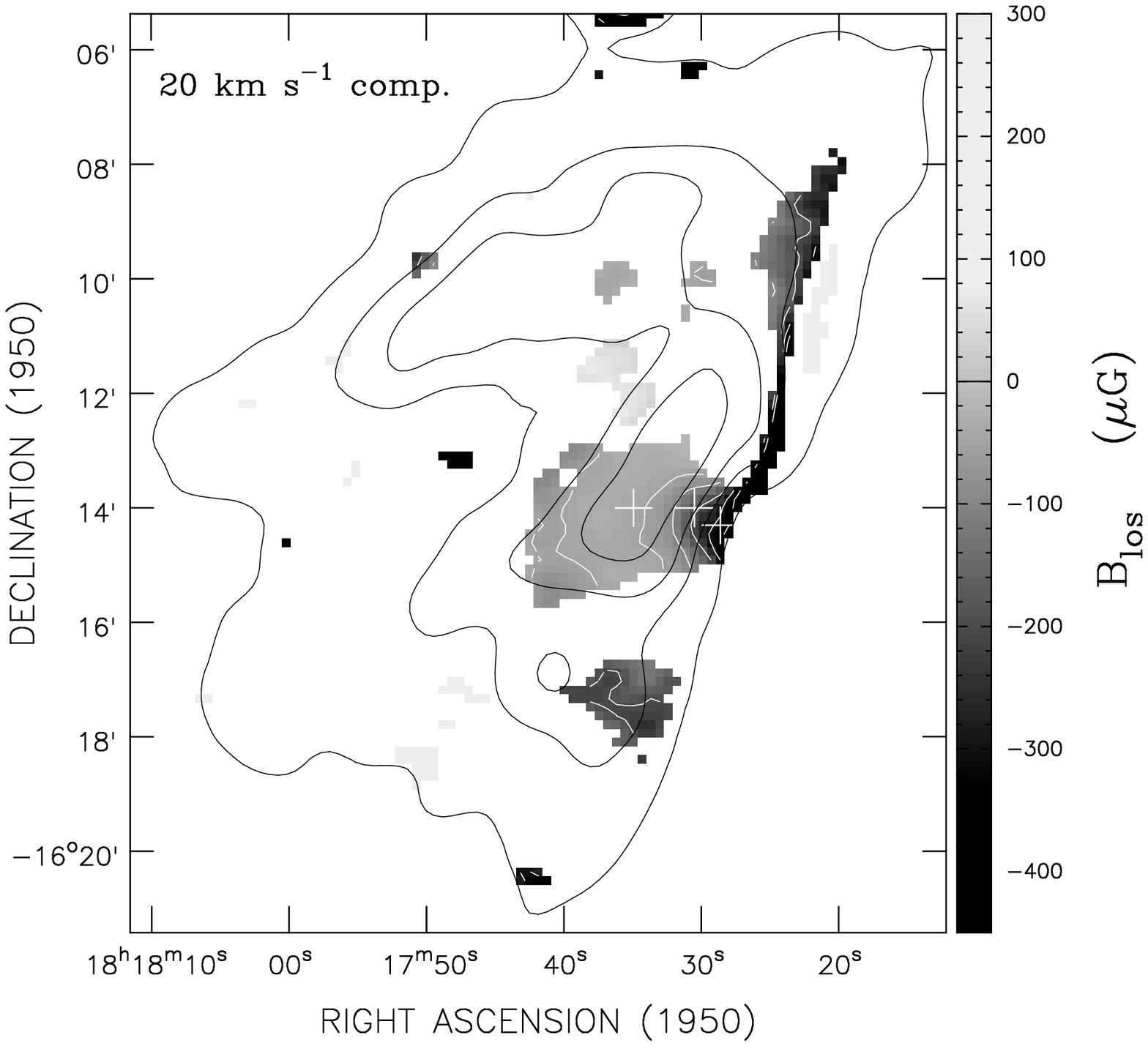}
\caption[Fig6.blos.eps]{Greyscale map of \Blos\/ for the 20 \kms\/ component with white
contours at $-$400, $-$300, $-$200, $-$100, and $-$50 \muG\/.  The 21 cm 1, 5, 10, and 20 Jy
beam$^{-1}$ continuum contours are shown in black.  The three white + symbols show the positions
of the profiles shown in Figs.  7 a, b, c \label{figu6}.}
\end{figure}

\placefigure{figu5}

Observations of M17 have been made for many astrophysically relevant molecular and atomic
transitions.  From these data it is known that M17 SW has a LSR velocity of 19 to 21 \kms\/
which corresponds to the velocity of one of our strongest \HI\/ absorption components.  Figure 5
shows an \HI\/ optical depth profile and several molecular emission profiles from
\markcite{stut88}Stutzki et al.  (1988) for a position at the \HII\/ region$-$molecular cloud
interface.  Note the similarity in the center velocities ($\sim$ 20 \kms\/) of each species.
This velocity coincidence strongly suggests that the 20 \kms\/ \HI\/ absorption component is
closely associated with the molecular cloud, most likely arising in the PDR at the \HII\/
region-molecular cloud interface.  Further evidence for this association comes from the steep
rise in the 20 \kms\/ \HI\/ optical depth toward the interface region (Fig.  4b).  Therefore,
values of \Blos\/ derived from the \HI\/ Zeeman effect in this component are of direct relevance
to the energetics of the molecular cloud and its immediate surroundings (see\S4.1.2).

\begin{figure}
\vspace{-3.5cm}
\epsfxsize=18.0cm \epsfbox{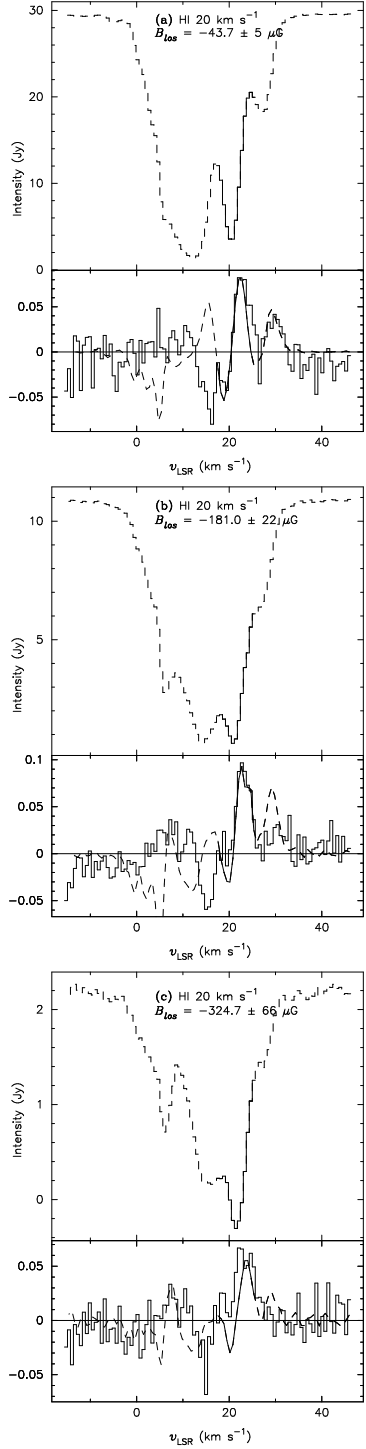}
\vspace{-2.0cm}
\figcaption[fig20prof.eps]{Representative fits at three positions for the 20 \kms\/ component
(17.0 to 24.6 \kms\/).  The upper panels show Stokes I profiles ({\em dashed histogram}), and
the bottom panels show Stokes V profiles ({\em solid histogram}) with the fitted derivative of
Stokes I shown as a smooth dashed curve.  In the upper and lower panels, the velocity range used
in the fit is denoted by solid or heavier lines.  The value of \Blos\/ fit for each position and
its calculated error are given at the top of each plot.  (a) Position $18^{\rm h}17^{\rm
m}35^{\rm s}$, ${-16}\arcdeg14\arcmin00\arcsec$.  (b) Position $18^{\rm h}17^{\rm m}30.6^{\rm
s}$, ${-16}\arcdeg14\arcmin00\arcsec$, or position (-60, -30) from 
Stutzki et al.  (1988).  (c) Position $18^{\rm h}17^{\rm m}28.5^{\rm s}$,
${-16}\arcdeg14\arcmin16\arcsec$, or position (-90, -45) from Stutzki et al.
(1988) \label{figu7}.}
\end{figure}

\placefigure{figu6}
\placefigure{figu7}

A significant magnetic field exists in the 20 \kms\/ component over much of the central region
of the southern bar.  The velocity range of the fit for this component is 17.0 to 24.6 \kms\/.
A distinct \HI\/ component at $\sim$ 23 \kms\/ is also visible toward the northwestern corner of
the source.  A component at this velocity was also detected at this velocity by \markcite{lad75}
Lada \& Chaisson (1975) in H$_2$CO.  However, detailed inspection of the \HI\/ data cube shows
that the channel range used to fit the 20 \kms\/ component was such that no magnetic field was
detected in regions where the 23 \kms\/ component is distinct.  Therefore, the \Blos\/ detected
in the velocity range 17 to 24.6 \kms\/ is only representative of the field in the 20 \kms\/
\HI\/ component.  A map of \Blos\/ for the 20 \kms\/ component is shown in Fig.  6, and sample
fits at representative pixels are shown in Figs.  7a, 7b, and 7c.  Negative values indicate that
\Blos\/ is toward the observer.  Values of \Blos\/ for this component (with sufficiently high
$S/N_B$ to pass our cutoff of 3) are centered near the continuum peak of the southern bar.  The
values for \Blos\/ start out at $\sim-$100 \muG\/ on the eastern side of the southern bar, pass
through a shallow saddle point and then rapidly rise to $\sim-$450 \muG\/ on the western side of
the map near the \HII\/ region$-$molecular cloud interface.  The maximum $S/N_B$ obtained for
this component is 8.5.  The apparently rapid change in direction of the \Blos\/ along the narrow
band in the northwest portion of the map may be real but is approaching the limit of our
sensitivity and will require future observation to confirm.

\begin{figure}
\vspace{-4.4cm}\hspace{1.0cm}
\epsfxsize=13.7cm \epsfbox{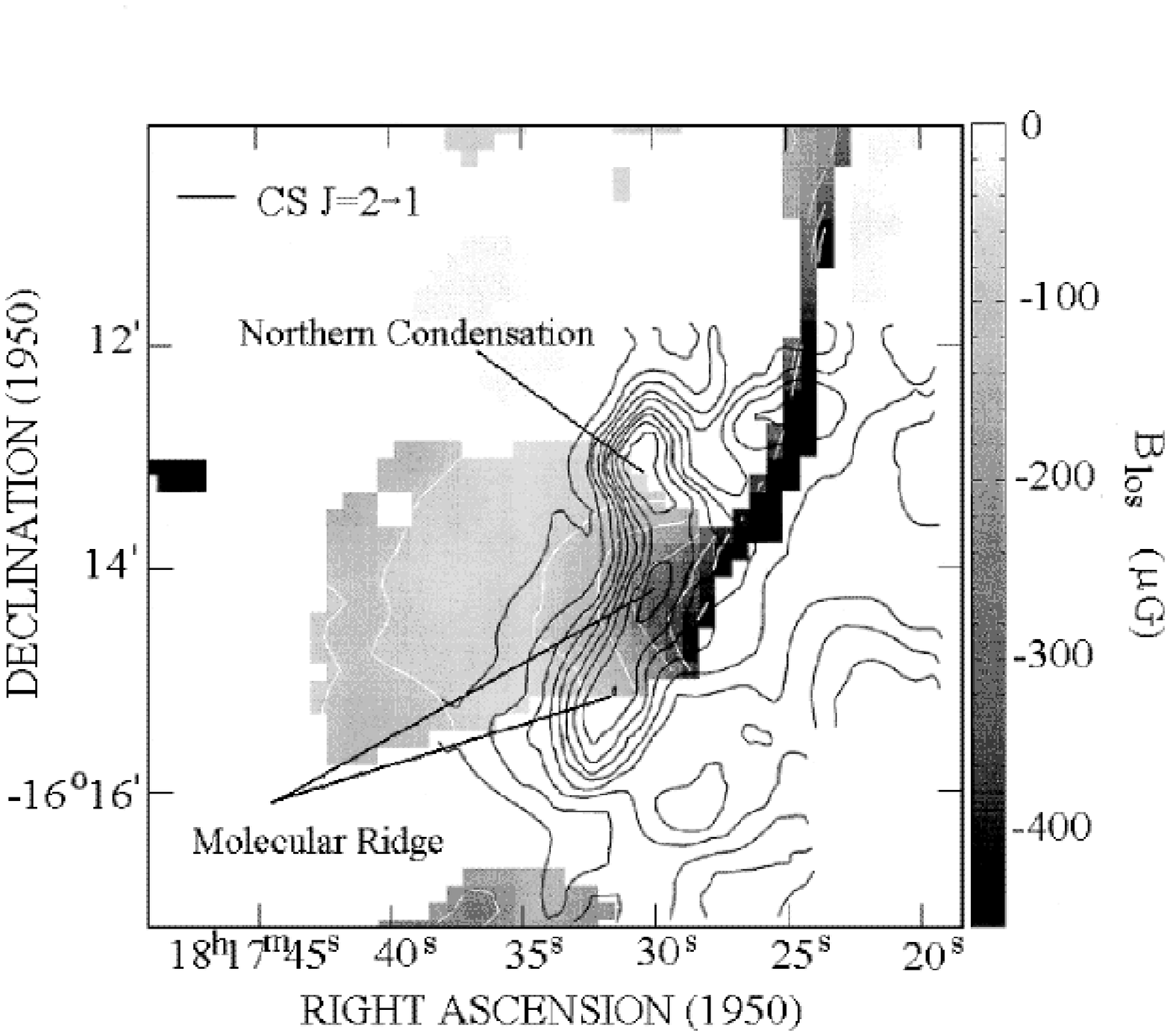}
\vspace{-0.2cm}
\caption[cryblos_cs2.eps]{Magnified greyscale of the \Blos\/ for the 20 \kms\/ component with
black CS $J=2\rightarrow~1$ contours from Wang et al.  (1993) superimposed.  The
white \Blos\/ contours are at $-$400, $-$300, $-$200, $-$100, and $-$50 \muG\/.  Notice the
spatial coincidence between the highest values of \Blos\/ and the molecular ridge traced by CS
emission \label{figu8}.}
\hspace{2.0cm}\hspace{2.0cm}
\epsfxsize=9.0cm \epsfbox{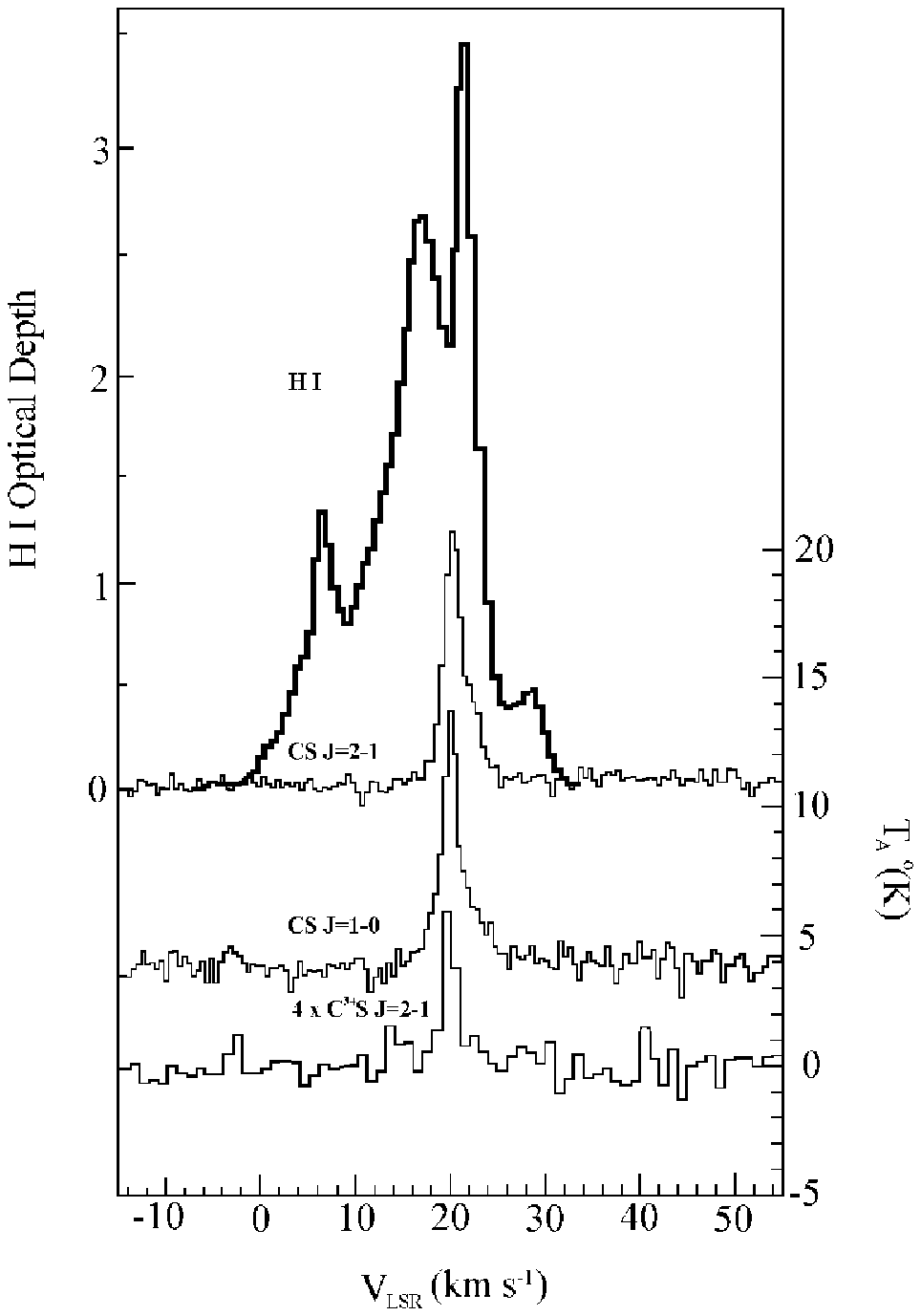}
\vspace{-1.4cm}
\figcaption[crywang5.eps]{An \HI\/ optical depth profile (present work) and emission spectra from
several CS transitions toward the molecular ridge ($18^{\rm h}17^{\rm m}32.4^{\rm s}$,
${-16}\arcdeg15\arcmin06\arcsec$) from Wang et al.  (1993).  See also Fig.8 \label{figu9}.}
\end{figure}

\placefigure{figu8}
\placefigure{figu9}

Note that the region of highest measured \Blos\/ coincides closely with the ridge of CS emission
observed by \markcite{wan}Wang et al.  (1993; Fig.  8).  Also, the 20 \kms\/ \HI\/ line in this
region is very narrow, and its width and center velocity correspond closely to those of the CS
lines (Fig.  9).  This velocity correspondence, like those of Fig.  5, provide further evidence
of a close association between the 20 \kms\/ \HI\/ gas and the molecular gas.  In addition, the
narrowness of the lines in this region seem to exclude the possibility of the ridge being
post-shock gas.  The nature of the \HI\/ gas at 20 \kms\/ and its effect on the morphology of
\Blos\/ will be discussed in \S4.1.2.
\newpage
\onecolumn
\newpage
\newpage
\twocolumn
\newpage

\subsubsection{Magnetic Fields Derived for the Blended 11$-$17 \kms\/ Components}

\begin{figure}
\vspace{-3.0cm}\hspace{2.0cm}
\epsfxsize=12.0cm \epsfbox{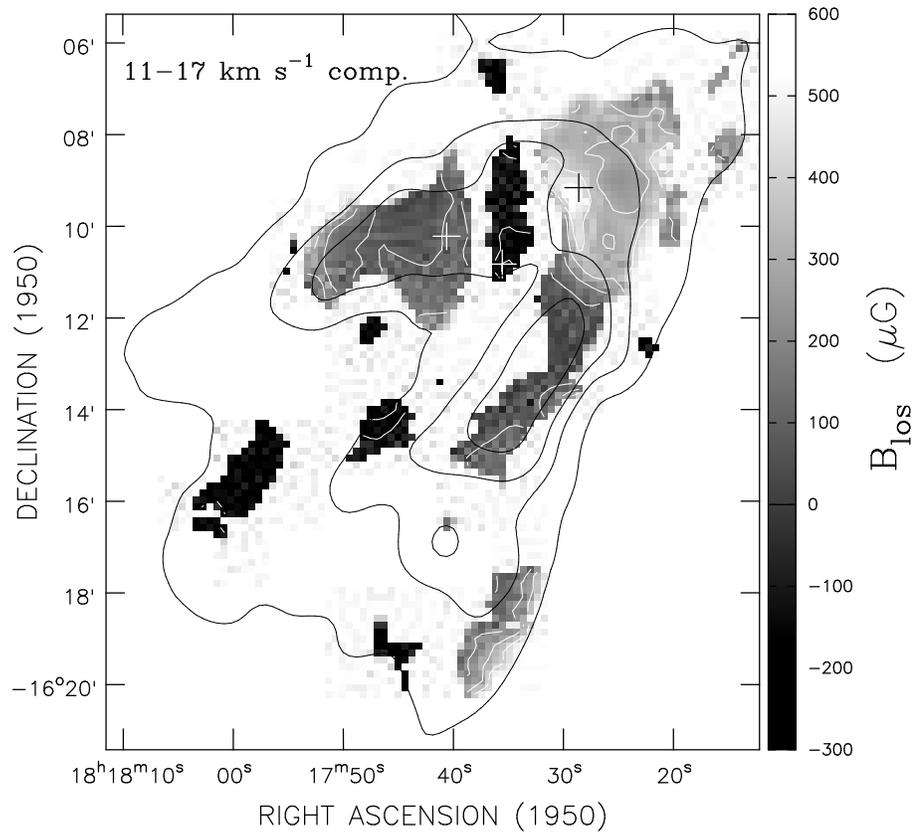}
\caption[Fig10.blos.eps]{Greyscale map of \Blos\/ for the 11$-$17 \kms\/ blended components
with white contours at $-$250, $-$150, 100, 200, 300, 400 and 500 \muG\/.  The 21 cm continuum
contours are shown in black at 1, 5, 10, and 20 Jy beam$^{-1}$.  The three + symbols show the
positions of the profiles shown in Fig.  11 a,b,c \label{figu10}.}
\end{figure}

\begin{figure}
\vspace{-3.5cm}
\epsfxsize=18.0cm \epsfbox{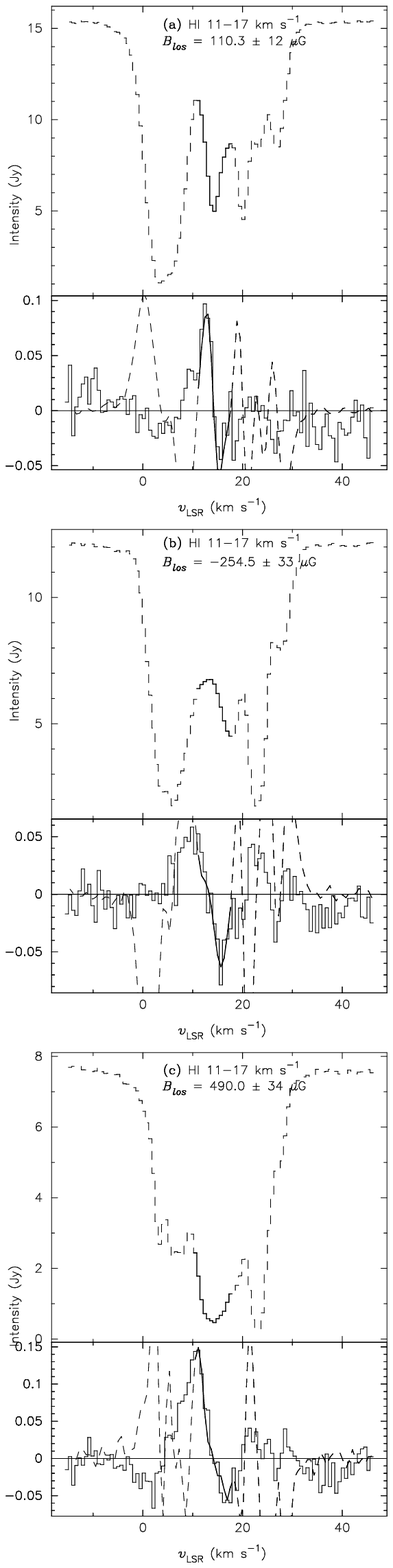}
\vspace{-2.0cm}
\caption[fig1117prof.eps]{Same as Fig.  7 a,b,c except the range of velocities used in the fit
for \Blos\/ was 11$-$17 \kms\/.  (a) Northeastern position $18^{\rm h}17^{\rm m}40.6^{\rm s}$,
${-16}\arcdeg10\arcmin15\arcsec$ (white + symbol in Fig.  10).  (b) North central position
$18^{\rm h}17^{\rm m}35.6^{\rm s}$, ${-16}\arcdeg10\arcmin51\arcsec$ (white + symbol in Fig.
10), and (c) Northwestern position $18^{\rm h}17^{\rm m}28.7^{\rm s}$,
${-16}\arcdeg09\arcmin12\arcsec$ (black + symbol in Fig.10).  Notice how the sign of the field
changes as you move from east to west across the map \label{figu11}.}
\end{figure}

\placefigure{figu10}
\placefigure{figu11}

Strong, primarily positive magnetic fields were also detected in blended velocity components
between 11 to 17 \kms\/.  A map of this magnetic field is shown in Fig.  10 and the fit to a few
representative pixels can be seen in Figs.  11a, 11b, and 11c.  There are four main regions of
significant \Blos\/ for this blended component which appear primarily in the northeastern, north
central and western parts of the map.  The center velocity of this component (defined by the
velocity at which the derivative changes sign) is somewhat different in each of these regions.
They range from 14.4 \kms\/ in the northeast (Fig.  11a), 13.7 \kms\/ in the central northern
feature (Fig.  11b), 14.0 \kms\/ in the northwest (Fig.  11c) and 12.4 \kms\/ in the southwest.
One interesting feature of the \Blos\/ for this component is that the direction of the field
changes from positive (away from the observer) to negative and back to positive across the map
from east to west (this trend can be seen in the fits shown in Figs.  11 a, b, c).  \Blos\/ in
the northwestern feature reaches values as high as +550 \muG\/ with $S/N_B$ as high as 14.

Evidence for molecular gas near this velocity has been observed in a number of species.  For
example \markcite{gre}Greaves, White, \& Williams (1992) reported a velocity component in this
range in their \tCO\/ $(3\rightarrow2)$ and \thCO\/ $(3\rightarrow2)$ spectra.  Their component
shifts from 12 \kms\/ in the southeast to 16 \kms\/ further to the southwest of their field of
view (the eastern portion of their maps overlaps the the western side of our maps).  The
positions where their low velocity component is centered at 12 \kms\/ corresponds to the
southwestern region of our magnetic field map where the \HI\/ line center is 12.4 \kms\/.
\markcite{rain}Rainey et al.  (1987) and \markcite{stut88}Stutzki et al.  (1988) also
identified emission in this velocity range in \tCO\/~$(3\rightarrow2)$ and
\tCO\/~$(2\rightarrow1)$, respectively.  \markcite{mas}Massi et al.  (1988) observed that two
of their seven NH$_3$ clumps near M17-UC1 have velocities of 15.5 \kms\/ and \markcite{stut88}
Stutzki et al.  (1988) observed 31 \CeO\/ clumps (out of 180) in the 11$-$17 \kms\/ velocity
range .

\placefigure{figu12}

In addition to molecular gas, there is also \CII\/ emission at a velocity of $\sim$ 17.8 \kms\/
(\markcite{bor}Boreiko et al.  1990).  \CII\/ is an important tracer of PDR's and should exist
in the same zone of a PDR as \HI\/ gas.  The precise spatial correlation between these two
atomic species for M17 is unknown.  However, the low resolution (3.\arcmin 4) map of \CII\/
obtained by \markcite{mat}Matsuhara et al.  (1989) does seem to follow the same morphology as
our summed N$_{HI}$/T$_s$ (see Fig.~4a).  A profile from one pixel at the \CII\/ peak from
\markcite{bor}Boreiko et al.  (1990; spatial resolution 43\arcsec\/) and \HI\/ are shown in
Fig.  12.  The \CII\/ profile appears to have a component at a lower velocity than their lowest
reported velocity of 17.8 \kms\/.  The authors noted that there may have been some contamination
from emission in the reference beam at this position.  The possible significance of a lower
velocity \CII\/ component will be discussed again in \S 4.1.3.

\subsubsection{Magnetic Field Detections at $\sim$ 3.4, 6.7, and 27 \kms\/}

Line of sight magnetic field detections were also made for three other velocity components.  A
significant \Blos\/ was measured in the velocity range $-$2.4 to 4.0 \kms\/ with the line center
at $\sim 3.4$ \kms\/.  This detection is interesting because the $S/N_B$ in the southeastern
region is as high as 10 and there is a rapid change in \Blos\/ from negative to positive values
along the northwestern border with M17 SW.  This abrupt change in field direction corresponds to
the behavior of the \Blos\/ measured in the 20 \kms\/ component in this same region.  The line
of sight fields in this component range from $\sim$ $-$200 to +100 \muG\/.

Detections were also made in the velocity range from 4.1 to 10.5 \kms\/ with a center velocity
of $\sim 6.7$ \kms\/ and the maximum 
\newpage
\onecolumn
\newpage
\twocolumn
\newpage
\noindent $S/N_B=6.4$. Reliable measurements of \Blos\/ in this
\HI\/ component range from $\sim$ $-$125 to +100\muG\/.  The regions of significant \Blos\/
measured for this component do not particularly correspond to the morphology of the 21 cm
continuum emission of M17.  In fact the highest fields and concentration of \HI\/ in this
velocity range are to the southeast.  One interesting possibility is that there is a connection
between \HI\/ measured in this velocity range and the 7 \kms\/ and the cold cloud observed by
\markcite{rieg}Riegel \& Crutcher (1972) at 21 cm over a large region toward the galactic
center including the direction of M17.  \markcite{cru84}Crutcher \& Lien (1984) estimate that
this cold cloud ($\sim$20 K) must be located within 150 pc of the sun.

A sparse \Blos\/ was also detected in the range 24.7 to 33.0 \kms\/ with a center velocity of
$\sim 27$ \kms\/ and maximum $S/N_B$=6.  The \Blos\/ in this \HI\/ component ranges from $\sim$
$-$200 to +100 \muG\/.  A component in this velocity range was also observed by \markcite{gre}
Greaves et al.  1992 in \tCO\/ $(3\rightarrow2)$ and \thCO\/ $(3\rightarrow2)$ emission.  The
ratio of these two species' intensity seems to indicate the presence of a distinct cloud at this
velocity.

\section{DISCUSSION}

\subsection{Proposed Origin of Magnetic Field Components}

\subsubsection{General Nature of the M17 Region}

\markcite{mei}Meixner et al.  (1992) suggest that the coincidence of part of their \OI\/ map
with the radio continuum emission indicates that hot ionized gas from the \HII\/ region is
carving a bowl into the molecular cloud to the southwest (M17 SW).  The symmetry axis of the
bowl lies in the plane of the sky, and the overlapping \OI\/ emission arises from a PDR on the
front and/or back sides of the bowl.  The bottom of the bowl (at the southwestern edge of the
southern bar) is the portion of the \HII\/ region-molecular cloud interface viewed edge$-$on.
Further evidence of a bowl$-$like morphology is the high visual extinction toward the southern
bar which presumably arises from the front side of the bowl.  \markcite{gul}Gull \& Balick
(1974) also suggest a bowl morphology based on the double peaks separated by $\sim$ 20 \kms\/ in
their H76$\alpha$ radio recombination line data.  (Also, see \markcite{gou}Goudis \& Meaburn
1976.)  These two recombination line velocity components are thought to arise from streaming
motions of newly-ionized gas at the front and back sides of the bowl.

In many ways, this M17 interface model resembles the interface region between the Orion nebula
and the molecular cloud behind it.  For Orion, the interface lies in the plane of the sky with
the ionizing stars directly in front of it.  For M17, the geometry is turned by 90$\arcdeg$.
The interface layer is perpendicular to the plane of the sky, with the ionizing stars offset to
the northeast.  The densest accumulation of ionized gas in M17 lies along the ``working
surface'' near the ionization front.  Therefore, the brightest region in the M17 continuum map,
the southern bar, does not coincide in the sky with the ionizing stars as it does for the Orion
nebula.

The clumpiness of the M17 interface has been observed in many molecular species (\S 1).  It
seems to closely resemble the scenario discussed by \markcite{bert96}Bertoldi\& Draine (1996),
whereby dense clumps are slowly photodissociated by a nearby \HII\/ region.  In this case the
\HII\/ region is partially embedded in (and partially obscured by) a bowl of clumpy neutral gas.
The \HI\/ gas associated with M17 can arise, in principle, anywhere in front of the source.  Our
data suggest that the 20 \kms\/ component originates primarily in the edge$-$on portion of the
bottom of the molecular bowl.  In this model the 20 \kms\/ \HI\/ arises from molecular gas that
been partially photodissociated but otherwise dynamically undisturbed (unshocked).  On the other
hand, the \HI\/ data for the 11$-$17 \kms\/ component supports the idea that it arises from
photodissociated, shocked gas from the front side of the molecular bowl which is streaming
toward us.  The association of different \HI\/ absorption components with shocked and unshocked
gas has also been suggested for S106 (Roberts et al.  1995) and W3 (Roberts et al.  1993).
Evidence supporting this picture for M17 is given in \S 4.1.2 and 4.1.3.

\subsubsection{The Nature of \HI\/ Gas at 20 \kms\/}

The velocity range of the 20 \kms\/ \HI\/ component (17 to 24.6 \kms\/) includes most of the
molecular gas in the M17 region (e.g \markcite{rain}Rainey et al.  1987 and \S 3.3.1).
Therefore, \HI\/ components in this range are likely to be closely associated with the bulk of
the M17 molecular gas.  Such an association is particularly clear toward the molecular ridge
(see Fig.  8) where the center velocity of the 20 \kms\/ \HI\/ component closely matches that of
a variety of molecular lines, including high density tracers such as CS (Figs.  5, 9, and 12).
This \HI\/ component is also relatively narrow (3-5 \kms\/), and its center velocity changes
little with position along the interface region.  Therefore, it most likely originates in
quiescent gas that is dynamically unaffected by the \HII\/ region.  Much of this quiescent gas
must lie in the edge-on \HII\/ region$-$M17 SW interface region at the bottom of the bowl
described by \markcite{mei}Meixner et al.  (1992).  Hence the concentration of 20 \kms\/ \HI\/
gas and high \Blos\/ ($\sim -450$ \muG\/) near the interface region (Figs.  3, 4b, 6, and 8).

\begin{figure}
\vspace{-3.8cm}\hspace{4.0cm}
\epsfxsize=6.8cm \epsfbox{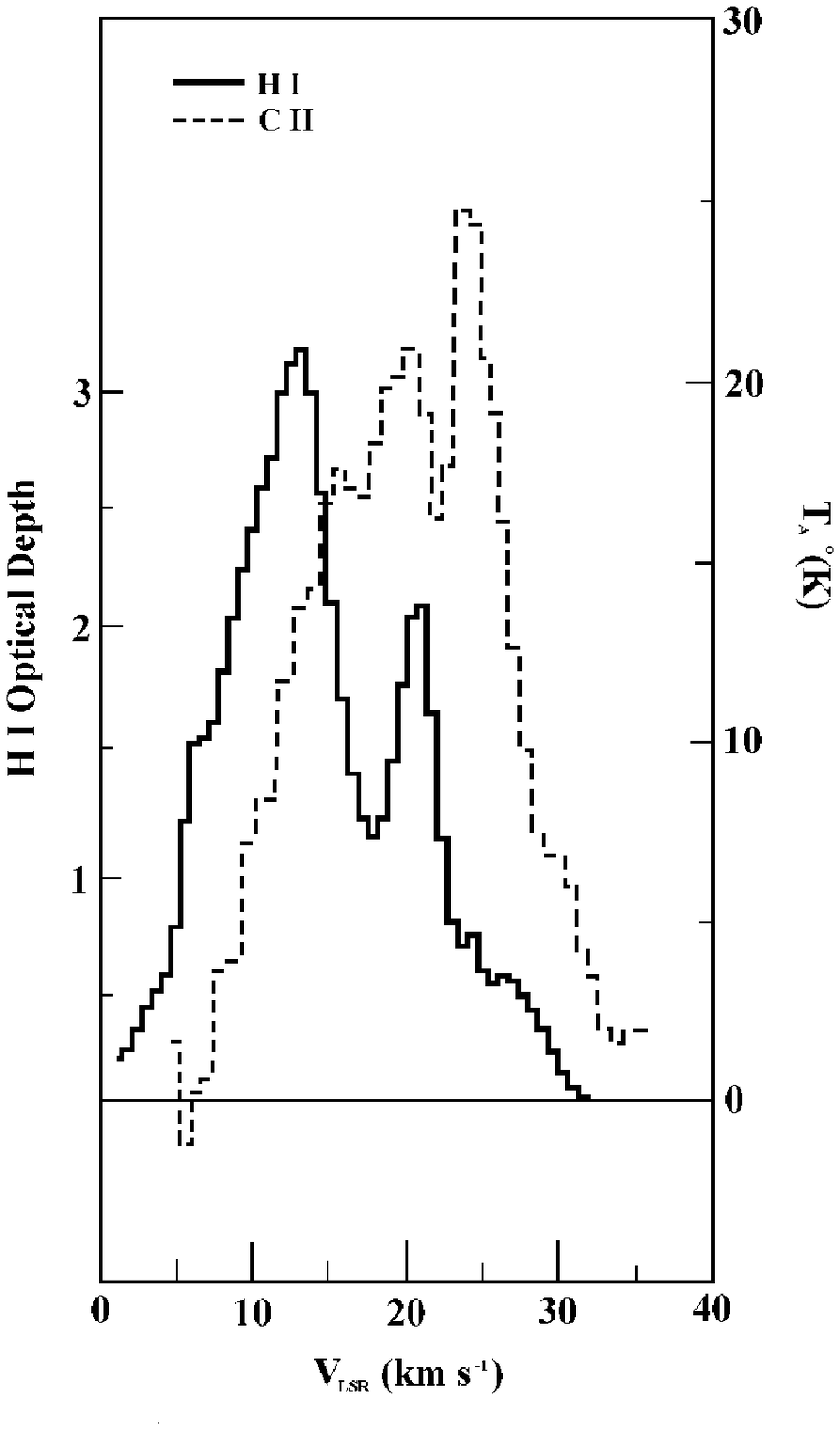}
\vspace{-0.7cm}
\caption[crycii2nd.eps]{An \HI\/ optical depth profile (present work) and a \CII\/ spectra
from Boreiko et al.  (1990) for the position of the Stutzki et
al.  (1988) \CII\/ ``peak'' $18^{\rm h}17^{\rm m}32.5^{\rm s}$,
${-16}\arcdeg13\arcmin42\arcsec$, or (-30, -15) \label{figu12}.}
\vspace{-0.4cm}\hspace{4.0cm}
\epsfxsize=8.0cm \epsfbox{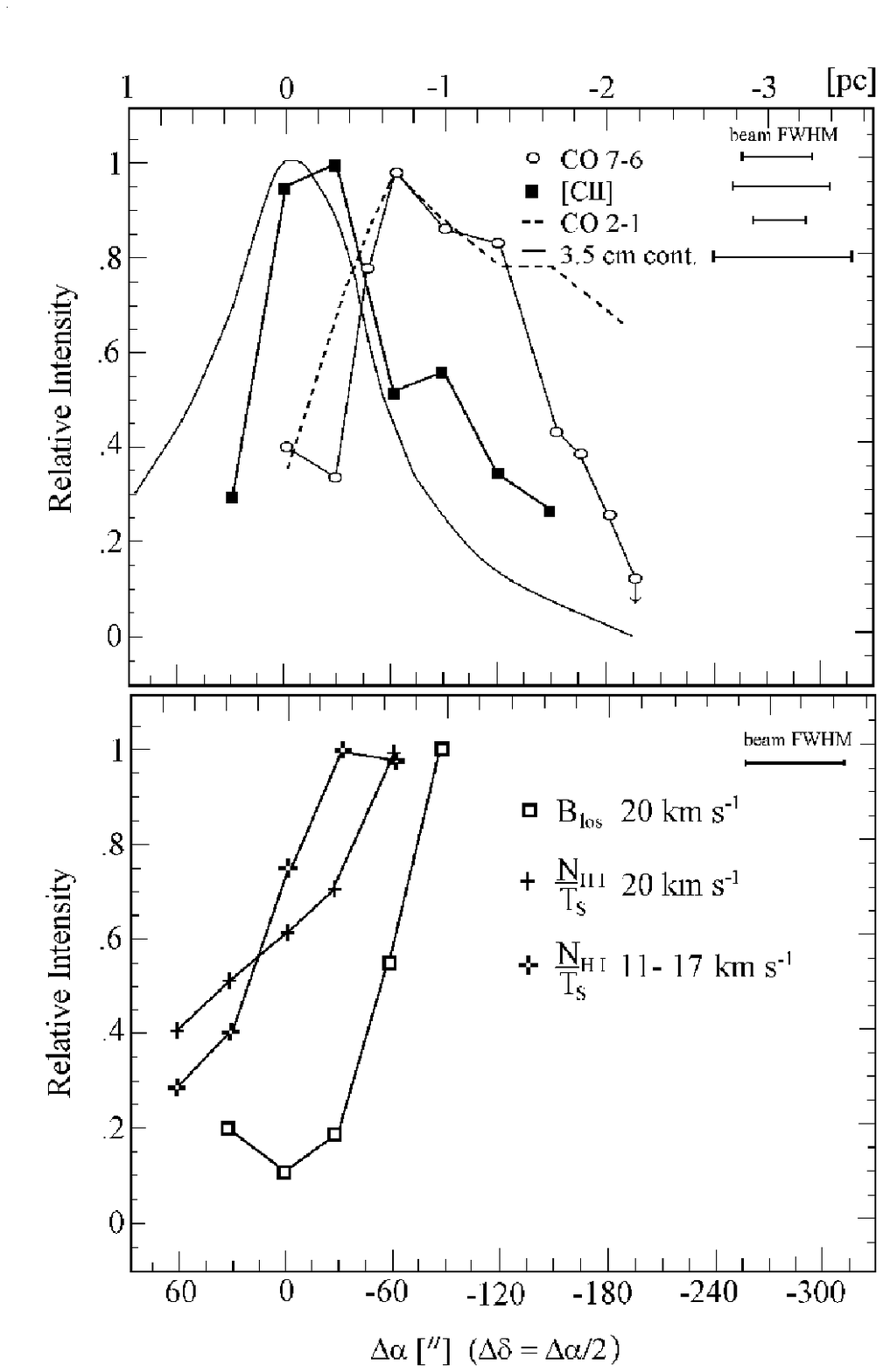}
\vspace{-2.9cm}
\caption[crytry6.eps]{ {\em Upper panel}~:  Distribution of the relative intensity of 3.5 cm
continuum, CO ($7\rightarrow 6$), CO ($2\rightarrow 1$) and \CII\/ across the \HII\/ region$-$
M17 SW interface following the NE-SW strip scan observed and plotted by 
Stutzki et al.  (1988).  {\em Lower panel}~:  Relative intensity of the 20 \kms\/ $N_{HI}/T_s$,
20 \kms\/ \Blos\/, and 11$-$17 \kms\/ $N_{HI}/T_s$ for the same positions as the upper panel.
The peak positions for the 20 \kms\/ $N_{HI}/T_s$ and 20 \kms\/ \Blos\/ do not necessarily
represent true maxima, due to the loss of continuum power as the strip is transversed to the
west.  Notice that the 11$-$17 \kms\/ $N_{HI}/T_s$ distribution along this strip is similar to
the distribution of \CII\/ emission (east of the continuum power drop-off), while the 20 \kms\/
$N_{HI}/T_s$ and 20 \kms\/ \Blos\/ peak further west \label{figu13}.}
\end{figure}

\placefigure{figu13}

Further indication of the association between \HI\/ and molecular gas comes from similarities in
their spatial distribution near the molecular ridge.  Fig.  13 (upper panel) shows the relative
intensity of several species along the NE-SW strip scan defined by \markcite{stut88}Stutzki et
al.  (1988).  The strip lies perpendicular to the interface region and cuts across it (ionizing
radiation propagates from left to right).  As shown in the figure, the increase in the 20 \kms\/
$N_{HI}/T_s$ and \Blos\/ (lower panel) matches well the increase in \tCO\/ line strengths to the
right of the \HII\/ region.  In fact, $N_{HI}/T_s$, \tCO\/$(2\rightarrow1)$, and
\tCO\/$(7\rightarrow6)$ peak {\em outside} of the \HII\/ region as expected for an interface
region viewed edge on.  Note that the actual peak in $N_{HI}/T_s$ for the 20 \kms\/ component is
not defined by our data since we cannot detect \HI\/ absorption beyond the continuum source.

Notice that the integrated \CII\/ emission, also shown in Fig.  13, peaks {\em inside} (to the
left of) the rise in the 20 \kms\/ $N_{HI}/T_s$.  In a simple planar PDR model, \CII\/ emission
coincides with \HI\/ and with molecular gas up to $A_v=2-4$ mag.  (See review
by\markcite{holl90}Hollenbach, 1990.)  However, the higher velocity resolution \CII\/ profiles
of \markcite{bor}Boreiko et al.  (1990) reveal the presence of several \CII\/ velocity
components.  An examination of these profiles suggests that the $\sim$ 20 \kms\/ \CII\/
component alone peaks further into the molecular cloud than the integrated \CII\/ emission
plotted in Fig.  13.  Also, note that the rise in the 11$-$17 \kms\/ $N_{HI}$/$T_s$, plotted in
Fig.  13 (lower panel), peaks to the left of the 20 \kms\/ $N_{HI}/T_s$ in a similar manner to
the integrated \CII\/ emission.  In fact, the contribution of lower velocity \CII\/ emission to
the integrated emission plotted in Fig.  13 may explain why it seems to peak to the left of the
20 \kms\/ $N_{HI}/T_s$.  Apparently, \HI\/ and \CII\/ gas is more broadly distributed along the
strip in Fig.  13 than is the gas at 20 \kms\/ alone.  This shows that absorbing \HI\/ gas lies
on the near side of the bowl, not just at the bottom.

The close association between \HI\/ gas near 20 \kms\/ and the molecular ridge region could
arise in at least two ways.  For one, the \HI\/ may be well mixed with and a minor constituent
of the molecular gas.  In such a case, the Zeeman effect near the ridge (Fig.  8) is directly
representative of magnetic fields in the molecular gas.  A second possibility is that the
absorbing \HI\/ near 20 \kms\/ is confined to a thin photodissociated shell about the molecular
ridge or to thin shells surrounding a multitude of molecular clumps within the ridge.  In this
case, the \HI\/ Zeeman effect is also representative of the magnetic field in the molecular gas
if the \HI\/ is photodissociated but dynamically undisturbed H$_2$.

In either case, the \HI\/ Zeeman effect near the 
\newpage
\onecolumn
\twocolumn
\newpage
\noindent molecular ridge must arise in relatively dense
gas if approximate equilibrium exists between the magnetic energy density and that associated
with non-thermal motions in the gas.  Such an equilibrium was suggested by \markcite{my88a}
\markcite{my88b}Myers \& Goodman (1988a, 1988b) on the basis of limited observational data.  In
such a case, the gas density is given by
\begin{equation} 
n_{eq} = \left[ \frac {B_{avg}}{0.4 \Delta v_{NT} }\right]^2 {\rm cm}^{-3}, 
\end{equation}
where $n_{eq}$ is the proton density, $B_{avg}$ is the average field strength (\muG\/), and
$\Delta v_{NT}$ is the FWHM contribution to the line width from non-thermal motions (\kms\/; see
also \S 4.3).  As an example we apply this equation to the position of Fig.  5 ($\sim 15\arcsec$
east of molecular ridge).  At this position, $\Delta v \approx 4$ \kms\/, and $B_{avg}$ \gtsim\/
200 \muG\/.  If $T_s=T_K=50$ K (as estimated by \markcite{bergin}Bergin et al.  1994 from
\tCO\/ observations), $\Delta v_{NT}$ = 3.7 \kms\/ and $n_{eq} \gtsim\/ 2 \times 10^4$ \cc\/.
This estimate of the local density in which the \HI\/ Zeeman effect arises is conservative since
the actual field strength $B_{avg}$ almost certainly exceeds \Blos\/ and $n_{eq} \propto
B_{avg}^2$.  This value for $n_{eq}$ based on \Blos\/ is also in excellent agreement with the
value of $n_{eq} \approx 6 \times 10^4$ \cc\/ calculated toward M17 SW by Goldsmith, Bergin, \&
Lis (1997) using \CeO\/ column densities and average line widths.  However, at the position
cited here, $N_{HI}/T_s$ in the 20 \kms\/ component is only 3 $\times 10^{19}$ \cmt\/K, so
$N_{HI} = 1.5 \times 10^{21}$ \cmt\/ for $T_s = $50 K.  A typical scale length for \HI\/
absorption features in the plane of the sky is of order 2 pc (3$\arcmin$ at 2.2 kpc).  If the
same scale length applies along the line of sight, then $n_{HI}$ $\approx$ 250 \cc\/.  That is,
if the \HI\/ Zeeman effect arises in gas at a density of order $n_{eq}$, then the \HI\/ could be
about only a 1\% constituent of largely molecular gas.  This possibility exists because the
$N_{H_2}$ estimated by \markcite{stut88}Stutzki et al.  (1988) and Goldsmith et al.  (1997)
near this direction ($\approx 2-8\times 10^{23}$ \cmt\/) is about 100 times higher than
$N_{HI}$.  Alternatively, \HI\/ might not be mixed with H$_2$ but confined to thin sheets or
very dense, thin atomic envelopes surrounding the multitude of molecular clumps which have been
observed in the interface region (see \S 1).  Note that this would still require the \HI\/ gas
to occupy about 1\% of the line of sight in the absorbing region.  These two cases are the same
ones cited above to explain the close association of \HI\/ gas with molecular gas in the ridge.

\subsubsection{The Nature of \HI\/ Gas at 11$-$17 \kms\/}

The morphology and velocity structure of gas between 11$-$17 \kms\/ suggest that it also lies
very close to the \HII\/ region on the near side of the bowl and that it has been shocked and
accelerated away from the ionized gas.  The 11$-$17 \kms\/ gas is primarily distributed in an
arc that partially encircles the northern and southern bars of the \HII\/ region (Figs.  3, 4c).
This arc of \HI\/ column density corresponds well with a similar arc structure apparent in
\tCO\/($3\rightarrow$2) emission mapped by \markcite{rain}Rainey et al.  (1987) in the velocity
range 14$-$20 \kms\/.  This morphology suggests a very close relationship between gas in the
11$-$17 \kms\/ velocity range and the \HII\/ region.  The gas is blue-shifted by 2-8 \kms\/
relative to the \HII\/ region center velocity of 19 \kms\/ (\markcite{jonc}Joncas \& Roy 1986).
Also, \HI\/ in this velocity range shows complex velocity structure with numerous components
appearing at various positions across the source (Figure 2, 11 a, b, c).  This complicated
velocity field and magnetic field reversal (see \S 3.3.2) suggest that gas in the 11$-$17 \kms\/
range lies in an interface zone that has been dynamically affected by interactions with the
\HII\/ region, unlike the 20 \kms\/ component discussed in \S 4.1.2.

Overall, the association between atomic and molecular gas in this velocity range is much less
clear than it is at 20 \kms\/ near the western interface region.  This could in part be due to
fact that few high resolution molecular observations have mapped the northern regions of M17.
However, judging from \tCO\/ $J=1\rightarrow$0, \tCO\/$J=3\rightarrow$2, and \thCO\/
observations (\markcite{lad76}Lada 1976; \markcite{rain}Rainey et al.  1987) which all show
components in this velocity range, the density of molecular material in this region is low
compared to the western interface.  Toward the northern bar in particular, where the 11$-$17
\kms\/ \HI\/ magnetic fields are high (see \S 3.3.2), there is not always a molecular
counterpart in the 11$-$17 \kms\/ velocity range.  In fact, for the positions shown in Figs.
11a and 11b there is no evidence of molecular gas.  Such a variable ratio of atomic to molecular
gas might be expected if gas in this velocity range has been subjected to a widely variable
degree of dissociation in the interface region.

Although molecular column densities are low in the 11$-$17 \kms\/ range, \HI\/ field strengths
are very high, reaching values of over +500 \muG\/ (Fig.  10).  Therefore, the assumption of
equilibrium between magnetic and non-thermal energies leads to conclusions similar to those for
gas near 20 \kms\/ (\S 4.1.2).  That is, $n_{eq}\gtsim\/10^4$ \cc\/, and $n_{HI}$ along the line
of sight is about two orders of magnitude less than $n_{eq}$.  Yet the scarcity of molecular gas
at positions of high field strength means that \HI\/ at these positions cannot exist as a minor
constituent in a largely molecular region.  If the \HI\/ alone is to exist at a density near
$10^4$ \cc\/, it must reside in one or more thin layers that occupy only about 1\% of the path
length through the \HI\/ absorbing region.  That is, the absorbing \HI\/ gas is distributed in a
very clumpy fashion along the line of sight.  This conclusion is consistent with numerous other
indications of a highly clumped medium in the M17 region (\S 1).  Most likely, the absorbing
\HI\/ gas is shocked, photodissociated gas lying in thin interface regions about ablating
neutral clumps.  The clumps, in turn, lie within mostly ionized gas, like the clumps modeled by
Bertoldi \& Draine (1996) or the clumps identified by \markcite{fel}Felli et al.  (1984) in
high resolution M17 radio continuum maps.  In these clumps, the pressure of the ionized gas
drives a shock front into the neutral medium.  Therefore, a rough equivalence between the
thermal energy density of the ionized gas and the magnetic energy density in the adjacent
compressed atomic region is expected.  Evidently, this is the case.  If the M17 ionized gas has
a temperature of 8000 K (\markcite{sub}Subrahmanyan \& Goss 1996) and a density of $10^4$ \cc\/
(\markcite{fel}Felli et al.  1984), then its thermal energy density is equivalent to the
magnetic energy density in the thin \HI\/ layers for B $\approx$ 500 \muG\/.

Several reasons exist to expect \HII\/ region driven motions of a few \kms\/ near M17.
Observationally, \markcite{rain}Rainey et al.  (1987) fit their \thCO\/ data on molecular
clumps near the southern bar to a geometric expansion model about Kleinemann's star (Fig.  1).
They obtain an expansion velocity of $\approx$ 11 \kms\/ about this star.  Radio and optical
recombination line observations often reveal split profiles shifted by about $\pm 8$ \kms\/ from
the \HII\/ region rest velocity (\markcite{gul}Gull \& Balick 1974; \markcite{jonc}Joncas \&
Roy 1986).  Theoretically, \markcite{bert96}Bertoldi\& Drain (1996) estimated the shock-induced
acceleration expected in the Orion nebula PDR, obtaining a value of 3 \kms\/.  This value is
somewhat sensitive to the ratio of post to pre-shock density, taken by \markcite{bert96}
Bertoldi\& Draine (1996) as 2.  If the ratio for M17 is 4, as suggested by the ratio of the
square of the \tCO\/ $(7\rightarrow 6)$ (5 \kms\/) line width (\markcite{har}Harris et al.
1987) to the square of the CS $(2\rightarrow 1)$ line width (2.5 \kms\/, \markcite{wan}Wang et
al.  1993), then the shock induced acceleration is increased to 6 \kms\/.  Also,
\markcite{bert89}Bertoldi) and \markcite{bert90}Bertoldi \& McKee (1990) discuss a rocket
effect upon ablating clumps of neutral gas exposed to ionizing radiation.  They estimate that
rocket acceleration should amount to a clump velocity of 5$-$10 \kms\/ under typical
circumstances.  These estimates of shock speed, agree quite well with the 2$-$8 \kms\/ blueshift
of the 11$-$17 \kms\/ componets.

\subsection{Comparison of 20 \kms\/ \Blos\/ Field Morphology with Polarimetry Studies}

A number of studies have been made of linear polarization toward M17.  These include
polarization at optical and near IR wavelengths from grain absorption and polarization in the
far IR from grain emission.  (See references cited by \markcite{dot}Dotson 1996.)  Such studies
reveal the position angles of the magnetic field in the plane of the sky ($B_{\bot}$), but they
yield little if any information about field strengths.  Therefore, they compliment Zeeman effect
studies, assuming the linear and circular polarization arise in the same regions.  In general,
field directions inferred from grain absorption do not agree with those inferred from grain
emission.  Very likely, this disprepancy results from the different regions sampled by each
technique, with grain emission data more likely representative of the field directions inside
M17 SW (\markcite{good}Goodman et al.  1995).

\begin{figure}
\epsfxsize=16.0cm \epsfbox{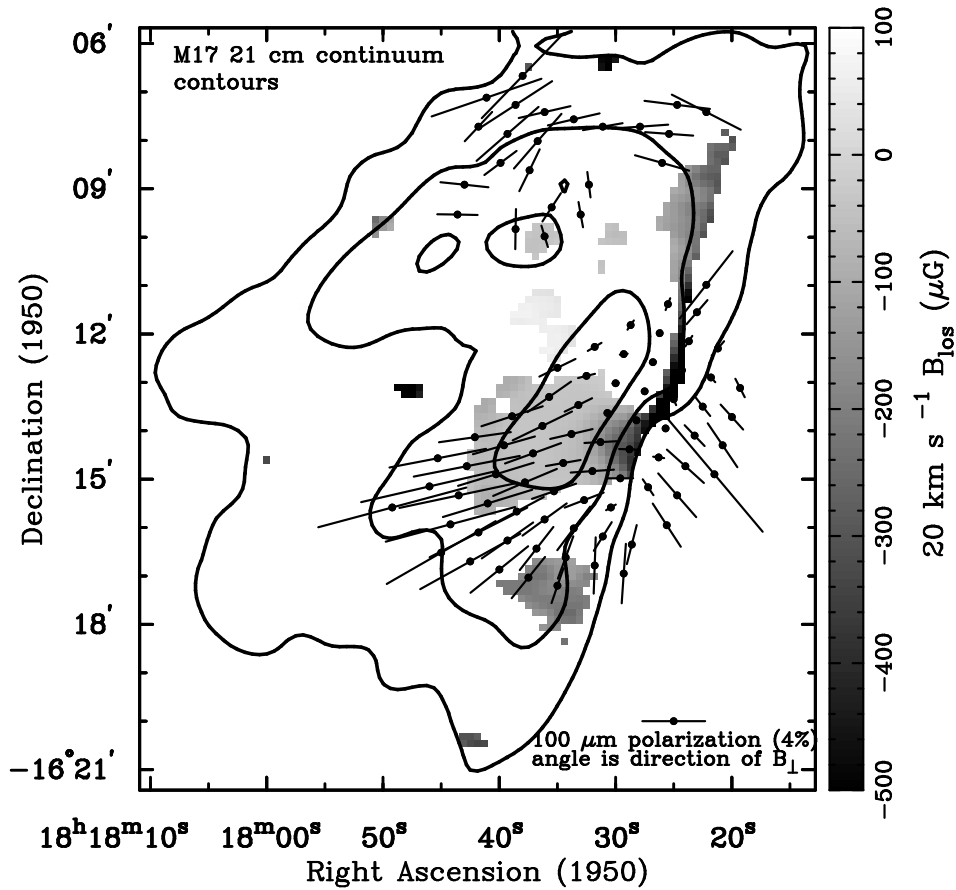}
\vspace{-4.0cm}
\caption[plotm17.eps]{Greyscale of the 20 \kms\/ \Blos\/ overlaid with the 1, 5, 15 Jy
beam$^{-1}$ contours from our 21 cm continuum, and the 100 \mum\/ linear polarization vectors
from Dotson (1996).  The position angle of the 100 \mum\/ polarization vectors
indicates the direction of $B{_\bot}$, which is perpendicular to the observed linear
polarization direction \label{figu14}.}
\end{figure}

\placefigure{figu14}

The study of polarized 100 \mum\/ dust emission by \markcite{dot}Dotson (1996) provides the
most comprehensive available data set for comparison with \HI\/ Zeeman effect results.  In
figure 14 we present a map of the position angle of the transverse magnetic field $B_{\bot}$
from \markcite{dot}Dotson (1996) overlaid on our $\sim$20 \kms\/ \Blos\/ grayscale.  The
resolution of the \markcite{dot}Dotson study (35$\arcsec$) is comparable to that of the \HI\/
data.  Also, the 20 \kms\/ \HI\/ component likely samples at least part of the molecular ridge
(\S 4.1.2) as does the 100 micron emission.  A principal conclusion of the \markcite{dot}Dotson
study is that the M17 magnetic field shows a significant degree of spatial coherence across M17.
The \HI\/ Zeeman data for the 20 \kms\/ component corroborates this result (Figs.  6, 8).
Moreover, the eight-fold {\em increase} in \Blos\/ from east to west toward the molecular ridge
(Fig.  8) is matched by a {\em decrease} in fractional linear polarization from about 4\% to
1\%.  Along this same line the field has a nearly constant position angle (predominantly
east-west).  Since \Blos\/ is perpendicular to both to the transverse magnetic field and the 100
\mum\/ linear polarization vectors, one might expect to see a correlation between regions of
maximum \Blos\/ and minimum 100 \mum\/ polarization.

One obvious interpretation of the IR and \HI\/ Zeeman data is that the magnetic field lines
curve into the line of sight as one approaches the molecular ridge from the east.  In such a
case, \Blos\/ may be a close approximation to the total field strength in the region of the
molecular ridge sampled by \HI\/.  Since the \HI\/ must lie in front of the \HII\/ region, the
field lines in this picture wrap around the \HII\/ region on the front side of the bowl,
becoming more aligned with the line of sight at the bottom of the bowl as they pass through the
molecular ridge.  This geometry is consistent with the inferences made by \markcite{dot}Dotson
from the linear polarization data alone.  The origin of the field curvature may lie in the
process of gravitational contraction that formed the molecular ridge, gathering an initially
uniform field along the line of sight into an hourglass shape with the ridge at its waist.  An
hourglass shaped field structure was also suggested for the W3 region by Roberts et al.  (1993),
although for this source, the axis of the hourglass is nearly in the plane of the sky rather
than along the line of sight.  Alternately, the present geometry of the M17 field may represent
the bending effects of the \HII\/ region as it expands into the molecular cloud deforming a
field originally along the line of sight into a bowl-like shape.

It should be noted that the occurrence of decreasing linear polarization toward the the 100
\mum\/ flux and column density maxima could result from effects other than a true minimum in the
transverse magnetic field component.  For example, optical depth effects, dust properties, grain
alignment efficiency, or some combination of these can cause a decrease in the linear
polarization independent of the magnetic field.  On the other hand, the magnetic field itself
could simply be more tangled in higher density regions (\markcite{jon}Jones, Klebe, \& Dickey
1992; \markcite{my91}Myers \& Goodman 1991).  In addition, the 760 \mum\/ polarimetry by
\markcite{val}Vall\'ee \& Bastien (1996) does not show a minimum in the linear polarization in
this region of the molecular ridge.  However, the 760 \mum\/ polarization arises from cooler
dust grains which could be spatially distinct along the line of sight from those emitting at 100
\mum\/.  Future polarimetry of this region at an intermediate wavelength from ISO and 
\newpage
\onecolumn
\twocolumn
\newpage
\noindent higher resolution \HI\/ Zeeman results may help to clarify this issue.

\subsection{Magnetic Energy vs. Turbulent and Gravitational Energy}

The magnetic field in a cloud can be separately described in terms of a static component \BS\/
and a time-dependent or wave component \Bw\/.  The static component connects the cloud to the
external medium and also determines the total magnetic flux through the cloud, while the wave
component is associated with MHD Alfv\'en waves in the cloud (e.g.  \markcite{Ar}Arons \& Max
1975).  Although the static component of the field can only provide support to the cloud
perpendicular to the field lines, the turbulent or wave component can provide 3-D support
(\markcite{mck95 }Mckee \& Zweibel 1995, also see references therein).  If a cloud is threaded
by $\Phi_B=\pi R^2$\BS\/, it can be completely supported perpendicular to \BS\/ if its mass does
not exceed the magnetic critical mass $M_{\Phi,crit}\approx 0.13G^{-{1/2}}\Phi_B$
(\markcite{mou76}Mouschovias \& Spitzer 1976).  This criterion leads directly to the
observationally relevant relation
\begin{equation}
B_{S,crit}\approx 5\times 10^{-21}N_{p}~\mu{\rm G}, 
\end{equation}
where $N_p$ is the average proton column density of the cloud.  If \BS\/ $>$ \Bscrit\/ (i.e.  $M
< M_{\Phi}$), then the cloud is supported by \BS\/, it is magnetically subcritical, and further
evolution of the cloud perpendicular to the field can only occur via ambipolar diffusion.  If
\BS\/ $<$ \Bscrit\/, (i.e.  $M > M_{\Phi}$), then \BS\/ cannot fully support the cloud, the
cloud is magnetically supercritical, and internal motions must supply additional support if the
cloud is stable.  In this case, further evolution of such a cloud will be controlled in part by
processes that create and dissipate internal motions.

As discussed above, the second type of support provided by magnetic fields arises from Alfv\'en
waves generated in a turbulent velocity field.  The fluctuating or wave component of the
magnetic field \Bw\/ organizes supersonic (but sub-Alfv\'enic) motions in the gas so that highly
dissipative shocks do not occur.  Theoretical studies suggest that equipartition between the
wave's kinetic and magnetic potential energy densities almost certainly exists in molecular
clouds (\markcite{zwe}Zweibel \& Mckee 1995, and references therein).  That is
$3/2\rho\sigma_{NT} =$ \Bw\/$^2 /8\pi$ where $\sigma_{NT}$ is the non-thermal component of the
velocity dispersion, which is assumed to arise exclusively from Alfv\'en waves.  These
assumptions lead to the equivalent of equation (2) for \Bw\/, that is,
\begin{equation}
B_W \approx 0.4 \Delta v_{NT} n_{p}^{1/2}~\mu {\rm G},
\end{equation}
where $\Delta v_{NT}$ is the non-thermal line width (FWHM) and $n_{p}$ is the proton density.
Since Zeeman effect measurements are generally sensitive to large scale fields (on the order of
the beam size), they are expected to be better indicators of \BS\/ than of \Bw\/.  Therefore,
the formal usefulness of comparing equation (4) for \Bw\/ to Zeeman observations (\Blos\/) is
not clear.  However, limited observational comparisons of Zeeman measurements with the \Bw\/
predicted from equation (4) show that there may also be approximate equipartition between the
wave's kinetic and static magnetic energy densities in molecular clouds; ie.
$\sigma_{NT}\approx 1/3v_A$, where $v_A^2=$\BS\/$^2/4\pi\rho$, or equivalently
\BS\/$\approx$\Bw\/ (\markcite{my88a} \markcite{my88b}Myers \& Goodman 1988a, 1988b; Bertoldi \&
Mckee 1992; \markcite{mou95}Mouschovias \& Psaltis 1995).  In any case it is worth noting that
\Bw\/ must be $\leq$ \BS\/ to prevent the formation of super-Alfv\'enic shock waves, which are
highly dissipative (\markcite{mck95}McKee \& Zweibel 1995).

We can assess the role of magnetic fields in the support of M17 SW from our \HI\/ Zeeman
measurements combined with other measurements of the clouds basic properties.  Goldsmith,
Bergin, \& Lis (1997) estimate that the mass of the M17 SW is $M_{core}=1.2 \times 10^4~{\rm
M_{\odot}}$ (including helium) on the basis of their \CeO\/ maps.  Using the geometric mean of
the extent of their M17 SW \CeO\/ maps (1.9 pc $\times$ 2.2 pc) $\sim 2$ pc as the $diameter$ of
the core region, and their integrated \CeO\/ column density, we estimate $<$\Np\/$>\approx 4
\times 10^{23}$ \cmt\/ and $<$\np\/$>\approx 6 \times 10^4$ \cc\/ (assumes filling factor of 1).
Although $\Delta v_{NT}$ is often estimated from the $\Delta v_{FWHM}$ of optically thin lines,
we believe a more accurate estimate for the internal motions of the M17 SW core as a whole comes
from the velocity dispersion of the center velocities of the clumps themselves.
\markcite{stut90}Stutzki \& G\"usten (1990) identified 180, $\sim 0.1$ pc \CeO\/ clumps in M17
SW and provide center velocities for each.  The dispersion of these center velocities is 2.7
\kms\/, equivalent to $\Delta v_{NT}$ = 6.3 \kms\/.  From these parameters and equations (3) and
(4), \Bscrit\/$\approx 2000$ \muG\/ and \Bw\/$\approx$ 620 \muG\/.  These estimates indicate
that (1) unless the static field is four or more times higher than our maximum values for
\Blos\/ (500 \muG\/), then M17 SW as a whole is magnetically supercritical and (2) the estimate
for \Bw\/ is in good agreement with our highest values for \Blos\/.  Therefore, the evolution of
the M17 SW core is not entirely controlled by \BS\/ and ambipolar diffusion, and internal
motions must play a role in its support and dynamics.  Note that these estimates are not
necessarily valid for individual clumps within the core (\markcite{bert92}Bertoldi\& McKee
1992).

Another perspective on cloud support can be gained from a simplified form of the virial theorem
that simultaneously takes into account magnetic effects from \BS\/ and \Bw\/ (but not external
pressure or rotation).  This form is
\begin{equation}
\mid {\cal W} \mid = {\cal M}_S + 3{\cal T}, 
\end{equation}
where \We\/ is the gravitational energy term, \Ms\/ is the magnetic energy associated with
\BS\/, and 3\Te\/ is the energy contribution from all internal motions including the magnetic
wave energy produced by \Bw\/ (See, for example, Mckee et al.  1993 for a more detailed
description of these terms.).  Note that 2T, in the non-magnetic virial equation, has been
replaced by 3T because equipartition between internal motions and Bw adds an additional term T
to the energetics of the cloud.  \markcite{mck93}McKee et al.  (1993) explain why virtually all
self-gravitating clouds must be close to dynamically critical since their internal pressures
would otherwise be comparable to external pressures, contrary to observation.  If M17~SW is
dynamically critical, that is, on the verge of collapse, then the virial equation above can be
rewritten in the form
\begin{equation}
B_S \approx B_{S,crit}[1 - (3{\cal T}/{\mid \cal W \mid})]^{1/2}.  
\end{equation}
Given the M17 SW cloud parameters estimated at the beginning of this section,
\Te\//$\mid$\We\/$\mid$ $\approx$ 0.3, so \BS\/ $\approx (1/3)$\Bscrit\/ $\approx 660$ \muG\/,
and $M_{\Phi}\approx 1/3M_{core}$.  Given the inevitable errors in estimating virial terms,
these values are quite uncertain.  However, they can be regarded as a possible equilibrium model
for the energetics of the M17 SW core.  In this model, internal motions in the core are
sub-virial since \Te\//\We\/ $<$ 1/2, and the cloud is magnetically supercritical.  That is, the
cloud is supported in part by internal motions, with the remainder of the support coming from
\BS\/.  Moreover, the magnitude of \BS\/ in this model is comparable to \Bw\/ (620 \muG\/), so
that the magnetic support provided by the static and wave components of the field are
comparable.  This suggests that \BS\/$\approx$\Bw\/ is a good assumption for M17 SW (see above).
Although our estimate for \BS\/ using this model is quite close to the highest measured values
of \Blos\/ ($\approx 500$ \muG\/), this agreement must be considered somewhat fortuitous since
estimates of \BS\/ are very sensitive to uncertainties in cloud parameters and, of course,
\Blos\/ is just one component of the field.

Apart from equations (3), (4), and (6), one other method has been used to estimate field
strengths in self-gravitating clouds.  Letting 2\Te\/$=\alpha \mid$\We\/$\mid$ we obtain the
following mass estimate:
\begin{equation}
M = \frac {5}{{\rm 8 ln 2}}~\left [\frac {\Delta v_{NT} ^2 R}{\alpha G }\right ],
\end{equation}
where $\alpha$ parametrizes the extent to which a magnetic field is needed to support the cloud,
ie.  $\alpha < 1$ for clouds in which the internal motions are sub-virial.  Substituting this
estimate for the cloud's mass into equation (4) we obtain
\begin{equation}
B_W = 15\alpha ^{-1/2}~\left [\frac {\Delta v_{NT} ^2}{R}\right]~ \mu{\rm G},  
\end{equation}
where $R$ is in pc and $\Delta v_{NT}$ is in \kms\/.  Note that this formulation is similar to
those found in \markcite{my88a} \markcite{my88b}Myers \& Goodman (1988a, 1988b),
\markcite{mou95}Mouschovias \& Psaltis (1995) (do not include $\alpha$), and \markcite{bert92}
Bertoldi\& McKee (1992) (using $\alpha$).  Comparing the mass computed from equation (7),
$M=\alpha^{-1}8.5 \times 10^3{\rm M_{\odot}}$, to the \CeO\/ mass of M17 SW ($1.2 \times
10^4~{\rm M_{\odot}}$), we find $\alpha=0.7$ and \Bw\/$\approx 740$ \muG\/, again, remarkably
similar to the peak measured value of \Blos\/.  The fact that the field strength predicted by
equation (8) is relatively well matched by \Blos\/ may be another indication that
\Bw\/$\approx$\BS\/ for most clouds (see above).

Although formulations similar to equation (8) have often been used to estimate magnetic fields
in the literature (see above), we consider equation (6) to yield a somewhat better field
estimate, at least in so far as comparison with Zeeman observations are concerned.  This is
because 1.)  if we assume $\alpha=1$ (\markcite{my88a} \markcite{my88b}Myers \& Goodman 1988a,
1988b), equation (8) is essentially the same as equation (4), both of which estimate \Bw\/.
There is evidence (this work, and references cited above) that \Bw\/$\approx$\BS\/, but this
assertion will require more observational evidence to confirm.  2.)  In addition, the assumption
that $\alpha$=1, excludes entirely the need for support from \BS\/, contrary to our
observational Zeeman evidence, ie.  \Blos\/$\approx 500$ \muG\/.  3.)  The inclusion of $\alpha$
in the equation solves this delima, but creates a new one, since the estimate of $\alpha$ only
describes the degree of under-virialization in the cloud, and does not distinguish between
support from \BS\/ and \Bw\/, while equation (8) is formally an estimate of \Bw\/ alone.  For
these reasons equation (6) may be the best method of estimating the magnetic field based on
cloud parameters for the purpose of comparison with Zeeman measurements even though all the
estimates found here (ie.  eq.  [4], eq.  [6], and eq.  [8]) are similar.

\section{SUMMARY AND CONCLUSIONS}

The \HI\/ absorption lines toward M17 show complicated profiles with 5 to 8 distinct velocity
components.  In general, the highest \HI\/ column densities are concentrated toward the \HII\/
region$-$M17 SW boundary.  This effect is particularly noticeable in the velocity range of the
20 \kms\/ \HI\/ component.  A lower limit to the maximum \HI\/ column density summed across the
whole \HI\/ velocity range is N$_{H_I}/T_s$= 1.0$\times 10^{20}$~\cmt\/.  Atomic and molecular
emission temperature measurements suggest that $T_s$ lies between 50 and 200 K.

A magnetic field has been detected at the same velocity as M17~SW ($v_{LSR}\sim 20$~\kms\/)
which peaks steeply toward the M17 \HII\/ region$-$M17 SW interface to values of \Blos\/= $-$450
\muG\/.  Analysis of this \HI\/ component's line width, along with its velocity and spatial
coincidence with molecular density tracers along the M17 SW molecular ridge suggest that it
originates in unshocked PDR gas.  Comparison of the proton density implied by the observed
\Blos\/ and the 20 \kms\/ \HI\/ density show that the \HI\/ gas is only a $\sim$ 1\% constituent
of the gas along the line of sight toward M17.  At the same time its distribution along the
\markcite{stut88}Stutzki et al.  (1988) strip scan across the \HII\/ region$-$M17 SW boundary
agrees well with that of molecular gas.  Therefore, we suggest that the 20 kms\/ \HI\/ gas lies
in thin, dense shells around the numerous molecular clumps which have been observed in M17 SW,
or it must be well mixed with fairly dense ($1\times 10^4$ \cc\/) interclump molecular gas.  In
addition, the region of maximum \Blos\/ for this component agrees spatially with a minimum in
100 \mum\/ polarimetry which samples the transverse magnetic field.

Another significant \Blos\/ was observed in the blended component at 11$-$17 \kms\/ with values
reaching $\sim +$ 550 \muG\/.  The morphology of high $N_{HI}$ for this component is quite
different from that at 20 \kms\/, with the greatest concentrations found further east and toward
the northern parts of the source.  From these components' morphology, line width, and evidence
that there is a significant amount of obscuring material on the front side of the M17 \HII\/
region, we suggest that the 11$-$17 \kms\/ components originate in shocked gas that is streaming
toward us along the line of sight.  Estimates of the shock speeds which can be produced by the
M17 \HII\/ region agree well with these components' blueshift (2$-$8 \kms\/) with respect to the
rest velocity of the region.  Like the 20 \kms\/ \HI\/ component, the 11$-$17 \kms\/ \HI\/ gas
can only comprise $\sim$ 1\% of the line of sight gas, and may be confined to photodissociated
\HI\/ shells surrounding molecular clumps close to the \HII\/ regions front side.

Using various virial arguments we have estimated that M17 SW is sub-virial (\Te\//\We\/$<$ 1/2)
and magnetically supercritical ($M_{core} > M_{\Phi}$).  In our model $\approx 1/3$ of M17 SW's
total support is provided by its static magnetic field \BS\/ ($\approx 660$ \muG\/), with the
rest of the support arising from internal motions including support from the wave component of
the field \Bw\/.  Estimates of the static (\BS\/) and wave (\Bw\/) components of the magnetic
field indicate that they are of approximately the same magnitude ($\approx 600 - 700$ \muG) and
agree well with our highest values of \Blos\/ ($\approx 500 $\muG\/).

Subsequent analysis of higher resolution VLA \HI\/ Zeeman effect data will give better
sensitivity to \Blos\/ if the field is tangled, and allow us to better differentiate separate
velocity components.  In addition, observations of the Zeeman effect in OH may also help
untangle the magnetic field data at this interface since it tends to trace somewhat higher
density gas.

\acknowledgments

C.  Brogan thanks NASA/EPSCoR for fellowship support through the Kentucky Space
Grant Consortium.  T.  Troland acknowledges support from NSF grant AST 94-19220, and R.
Crutcher acknowledges support from NSF grant AST 94-19227.

\newpage

\begin{deluxetable}{lrrrrcrrrrr}
\tablewidth{33pc}
\tablecaption{Observational Parameters of 21 cm HI Zeeman Observation of M17}
\tablehead{
\colhead{Parameter}           & \colhead{Value}}      
\startdata
Frequency & 1420 MHz \nl
Observing date & 1996 June 2 \nl
Total observing time & 8 hr \nl
Primary beam HPBW & 30\arcmin \nl
Synthesized beam HPBW & 61.0\arcsec $\times$ 44.5\arcsec \nl
Phase and pointing center(B1950) & $18{\rm^h}17{\rm^m}35.0{\rm^s}$, 
${-16}\arcdeg$14\arcmin00\arcsec \nl
Frequency channels per polarization & 256 \nl
Total bandwidth & 781.25 kHz (165.05 \kms\/) \nl
Velocity coverage & -67.5 to +97.5 \kms\/ \nl
Frequency resolution & 3.052 kHz (0.64 \kms\/) \nl
Rms noise in line channels & 40 mJy beam$^{-1}$ \nl
Rms noise in continuum & 80 mJy beam$^{-1}$ \nl
$S_{\nu}$ to $T_b$ & 1 mJy beam$^{-1}$ = 0.25 K \nl
Angular to linear $scale^a$ & 1\arcmin = 0.6 pc \nl
Position of 21 cm continuum peak & $18{\rm^h}17{\rm^m}32.5{\rm^s}$,
${-16}\arcdeg$13\arcmin06\arcsec \nl
\tablenotetext{a} {Assumes distance of 2.2 kpc.}
\label{tab1}

\enddata
\end{deluxetable}

\end{document}